%% file: Report.tex
\documentclass[12pt, a4paper]{report}
\usepackage[a4paper, total={6.25in, 8.25in}]{geometry} 

\input{Packages.tex}
\usepackage{multicol}
\hypersetup{pdftitle = Project Report, pdfauthor = {First Last}, pdfstartview=FitH, pdfkeywords = essay, pdfpagemode = FullScreen, colorlinks, anchorcolor = black, citecolor = blue, urlcolor = blue, filecolor = green, linkcolor = blue, plainpages = false}
\fancyheadoffset{8pt}
\fancyfootoffset{8pt}

\input{Glossary}

\date{September 2024}

\title{A Few Shot Learning Scheme for Quantum Natural Language Processing}
\author{\\ \Large{Juan Pablo Rubio Perez}
\\ Supervisors: Mehrnoosh Sadrzadeh
\\ Faculty of Maths and Physics
\\ Department of Physics and Astronomy
\\ 
\\
\\ University College London
\\
A Project Report Presented in Partial Fulfillment of the Degree \\ \textit{MSc in Quantum Technologies}
\\ \\
}
\begin{document}

\AddToShipoutPictureBG*{%
  \AtPageUpperLeft{%
    \raisebox{-\height}{%
      \includegraphics[width=\paperwidth]{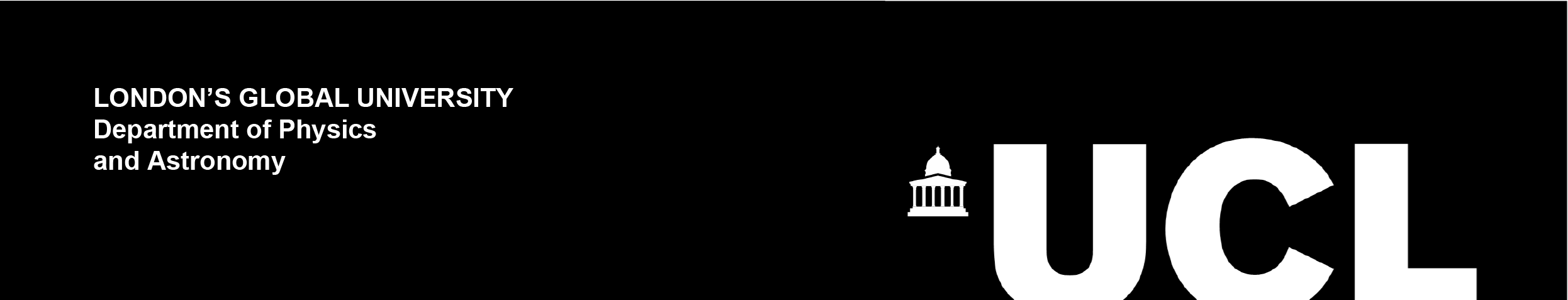}%
    }}
}
\AddToShipoutPicture*{%
      \parbox[t][\paperheight][t]{\paperwidth}{%
          \includegraphics[width=1.2\paperwidth]{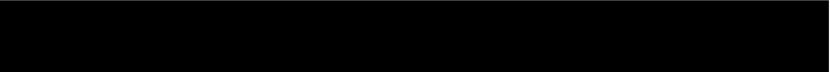}
      }}

\thispagestyle{headings}
\maketitle
\FloatBarrier
\pagenumbering{roman}

\thispagestyle{empty}
\begin{abstract}
\input{0_Abstract}

\end{abstract}
\newpage
\thispagestyle{empty}
\begin{center}
A mi pap\aa, que se qued\oo \vspace{1pt} en el camino y s\ee \vspace{1pt} que se alegrar\ii a de ver a d\oo nde he llegado y lo que me falta por recorrer, a mi mam\aa \vspace{1pt} y mi hermano por el apoyo que nunca se detuvo, y a todos los que quiero que de nombrarlos uno por uno no quedar\ii a m\aa s espacio para escribir y pecar\ii a de omisi\oo n.
\end{center}

\newpage
\thispagestyle{empty}
\vspace*{\fill}
\begin{center}
Copyright \copyright  \thinspace 2024 by Juan Pablo Rubio Perez \\ All Rights Reserved
\end{center}
\vspace*{\fill}
\newpage
\thispagestyle{empty}
\epigraph{But what...is it good for.}{--- \textup{\textit{Engineer at IBM, 1968, commenting on the microchip}}}

\epigraph{Radio has no future. Heavier-than-air flying machines are impossible. X-rays will prove to be a hoax.}{--- \textup{\textit{William Thomson, Lord Kelvin, 1899}}}

\epigraph{Nature isn't classical, dammit, and if you want to make a simulation of Nature, you'd better make it quantum mechanical, and by golly it's a wonderful problem because it doesn't look so easy.}{--- \textup{\textit{Richard Feynman, 1981}}}

\thispagestyle{empty}

\thispagestyle{empty}
\chapter*{Declaration}
I, Juan Pablo Rubio Perez,  declare that the thesis has been composed by myself and that the work has not be submitted for any other degree or professional qualification. I confirm that the work submitted is my own, except where work which has formed part of jointly-authored publications has been included and referenced. The report may be freely copied and distributed provided the source is explicitly acknowledged. \\
 \begin{figure}[H]
 \includegraphics[width=0.3\linewidth]{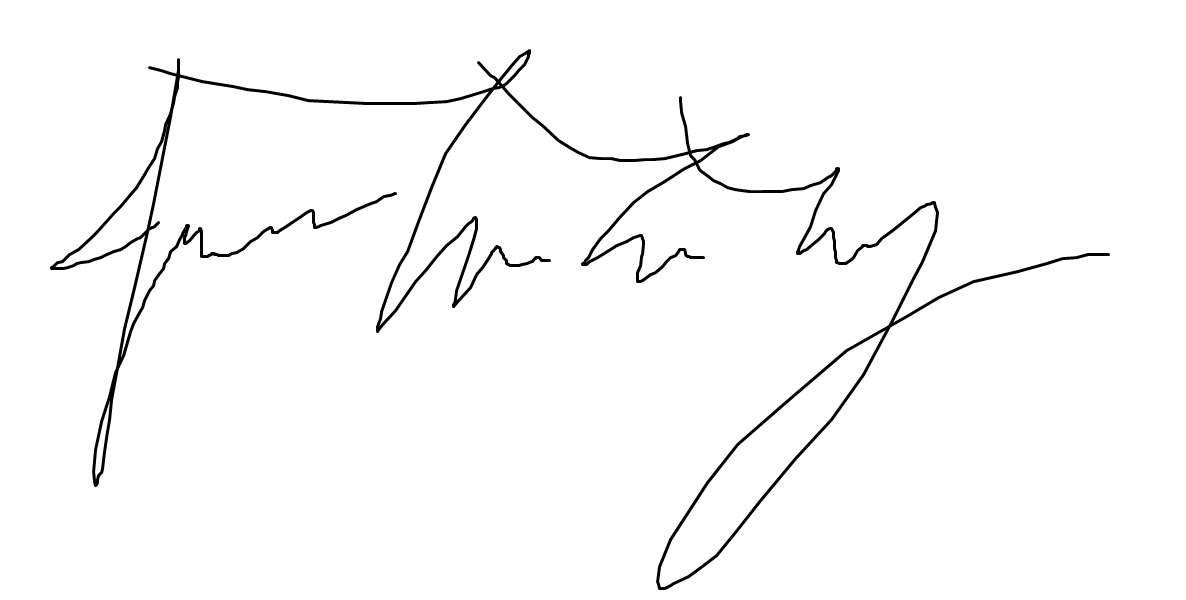}
 \end{figure}
\vspace{-2cm}
\noindent\begin{tabular}{ll}
 & 23/08/24 \\
\makebox[2.5in]{\hrulefill} & \makebox[2.5in]{\hrulefill}\\
\textit{Signature} & \textit{Date}\\
\end{tabular}

\tableofcontents

\thispagestyle{plain}
\listoffigures

\chapter*{List of Abbreviations}
\begin{sortedlist} 
  \sortitem{NISQ:    Noisy Intermediate Scale Quantum}
  \sortitem{QPC:    Quantum Parametrised Circuits}
  \sortitem{HQC:    Hybrid Quantum-Classical}
  \sortitem{QML:    Quantum Machine Learning}
  \sortitem{QNLP:    Quantum Natural Language Processing}
  \sortitem{QNLP:    Natural Language Processing}
  \sortitem{FSL:    Few Shot Learning}
  \sortitem{OOV:    Out of Vocabulary}
  \sortitem{VQE:    Variational Quantum Eigensolver}
  \sortitem{HEA:    Hardware Efficient Ansatz}
  \sortitem{QPE:    Quantum Phase Estimation}
  \sortitem{QFT:    Quantum Fourier Transform}
  \sortitem{qGAN:    Quantum Generative Adversarial Networks}
  \sortitem{QCNN:    Quantum Convolutional Neural Networks}
  \sortitem{BERT:    Bidirectional Encoder Representations from Transformers}
  \sortitem{PQE:    Pre Quantum Embeddings}
  \sortitem{BoW:    Bag of Words}
  \sortitem{FSL:    Few Shot Learning}
  \sortitem{AAE:    Approximate Amplitude Encoding}
  \sortitem{PQA:    Practical Quantum Advantage}
  \sortitem{PrPQA:    Provable Quantum Advantage}
  \sortitem{RPQA:    Robust Practical Quantum Advantage}

  \end{sortedlist}
\input{1_Introduction}
\input{2_Scheme}
\input{3_Theory}

\input{4_Problem}
\input{5_Framework}
\input{6_Experiment}

\input{7_ResandCon}

\renewcommand{\bibname}{References}
\bibliographystyle{ieeetr}
\bibliography{bibliography.bib}
\begin{appendices}
\chapter{Source Code} \label{System Requirements}
Source code for the datasets (both original and extended), the trained neural networks, and the code used can be found in: \newline \url{https://github.com/juanrubiop/QFsl}. \newline \\

\input{app_QuantumSupremacy}
\input{app_bias}
\input{app_aae}

\end{appendices}


\end{document}

%% file: Packages.tex
\usepackage[T1]{fontenc}
\usepackage[english]{babel}
\usepackage{siunitx}
\usepackage{graphicx}
\usepackage{tipa} 
\usepackage{graphics} 
\usepackage{amsmath}
\usepackage{latexsym}
\usepackage{amscd}
\usepackage{mathrsfs} 
\usepackage{relsize}
\usepackage{delarray}

\usepackage{graphicx} 

\usepackage{datatool}

\usepackage[table,xcdraw]{xcolor} 

\usepackage[nottoc]{tocbibind} 

\usepackage{glossaries}

\usepackage{pstricks}

\usepackage{theorem}
\usepackage{changepage}
\usepackage{euscript}
\usepackage{textcomp}
\usepackage{esvect}
\usepackage{parskip}
\usepackage{placeins}
\usepackage{subfig}
\usepackage{subcaption}
\usepackage{array}
\usepackage{delarray}
\usepackage{stmaryrd}
\usepackage{fancyhdr}
\usepackage{graphpap}
\usepackage{makeidx}
\usepackage{enumerate}
\usepackage{esint}
\usepackage{datetime}
\usepackage{caption}
\usepackage{smartdiagram}
\usesmartdiagramlibrary{additions}
\usepackage{abstract}
\usepackage{color}
\definecolor{igreen}{rgb}{0.0, 0.56, 0.0}
\usepackage{xcolor, colortbl}
\colorlet{gred}{-red!75!green!65!}
\colorlet{mamber}{-red!75!green!15!blue!50!}
\colorlet{grown}{-red!75!blue!20!green}
\colorlet{bled}{-red!85!blue!40!green!45!}
\colorlet{waters}{cyan!25} 
\colorlet{water}{cyan!25!green!20!} 
\definecolor{grin}{HTML}{00F9DE}
\usepackage{rotating}

\usepackage{arydshln}
\usepackage{booktabs}
\usepackage{graphicx}
\usepackage{tikz}
\usepackage{tikz-3dplot}
\usetikzlibrary{
arrows,
arrows.meta,
automata,
backgrounds,
calc,
decorations,
decorations.pathmorphing,
decorations.pathreplacing,
decorations.fractals,
external,
fit,
matrix,
petri,
positioning,
shadows,
shapes,
shapes.multipart,
topaths,
intersections
}
\usepackage{eso-pic}
\def\ba{\begin{array}}
\def\ea{\end{array}}
\def\beann{\begin{eqnarray*}}
\def\eeann{\end{eqnarray*}}
\def\bea{\begin{eqnarray}}
\def\eea{\end{eqnarray}}

\makeatletter
\newlength\qvec@height
\newlength\qvec@depth
\newlength\qvec@width
\newcommand{\qvec}[2][]{
    \settoheight{\qvec@height}{$#2$}
    \settodepth{\qvec@depth}{$#2$}
    \settowidth{\qvec@width}{$#2$}
  \def\qvec@arg{#1}
  \raisebox{.2ex}{\raisebox{\qvec@height}{\rlap{%
    \kern.05em
    \begin{tikzpicture}[scale=1,shorten >=-3pt,shorten <=-3pt]
    \pgfsetroundcap
    \coordinate (Stx) at (.05em,0) ;
		\coordinate (Arx) at (\qvec@width-.05em,0) ;
    \draw[->](Stx) to[bend left] (Arx);
    \end{tikzpicture}
  }}}
  #2
}
\makeatother
\makeatletter
\newlength\pvec@height
\newlength\pvec@depth
\newlength\pvec@width
\newcommand{\pvec}[2][]{
    \settoheight{\pvec@height}{$#2$}
    \settodepth{\pvec@depth}{$#2$}
    \settowidth{\pvec@width}{$#2$}
  \def\pvec@arg{#1}
  \raisebox{.2ex}{\raisebox{\pvec@height}{\rlap{%
    \kern.05em
    \begin{tikzpicture}[scale=1,shorten >=-3pt,shorten <=-3pt]
    \pgfsetroundcap
    \coordinate (Stx) at (.05em,0) ;
		\coordinate (Arx) at (\pvec@width-.05em,0) ;
    \draw[->](Stx) to[bend right] (Arx);
    \end{tikzpicture}
  }}}
  #2
}
\makeatother
\makeatletter
\newlength\vvec@height%
\newlength\vvec@depth%
\newlength\vvec@width%
\newcommand{\vvec}[2][]{%
  \ifmmode%
    \settoheight{\vvec@height}{$#2$}%
    \settodepth{\vvec@depth}{$#2$}%
    \settowidth{\vvec@width}{$#2$}%
  \else 
    \settoheight{\vvec@height}{#2}%
    \settodepth{\vvec@depth}{#2}%
    \settowidth{\vvec@width}{#2}%
  \fi%
  \def\vvec@arg{#1}%
  \def\vvec@dd{:}%
  \def\vvec@d{.}%
  \raisebox{.2ex}{\raisebox{\vvec@height}{\rlap{%
    \kern.05em%
    \begin{tikzpicture}[scale=1]
    \pgfsetroundcap
    \draw (.05em,0)--(\vvec@width-.05em,0);
    \draw (\vvec@width-.05em,0)--(\vvec@width-.15em, .075em);
    \draw (\vvec@width-.05em,0)--(\vvec@width-.15em,-.075em);
    \ifx\vvec@arg\vvec@d%
      \fill(\vvec@width*.45,.5ex) circle (.5pt);%
    \else\ifx\vvec@arg\vvec@dd%
      \fill(\vvec@width*.30,.5ex) circle (.5pt);%
      \fill(\vvec@width*.65,.5ex) circle (.5pt);%
    \fi\fi%
    \end{tikzpicture}%
  }}}%
  #2%
}
\makeatother
\def\ba{\begin{array}}
\def\ea{\end{array}}
\def\beann{\begin{eqnarray*}}
\def\eeann{\end{eqnarray*}}
\def\bea{\begin{eqnarray}}
\def\eea{\end{eqnarray}}

\usepackage{titlesec}
\usepackage{multirow}
\usepackage{hyperref}
\usepackage{lipsum}
\usepackage{tikz-cd}
\usepackage{float}
\usepackage{titling}
\usepackage{epigraph}
\usepackage[title, titletoc]{appendix}
\titleformat{\chapter}{\normalfont\LARGE}{\thechapter\,$\vert$}{20pt}{\LARGE}{\setcounter{chapter}{0}}
\titlespacing*{\chapter}{0pt}{-70pt}{40pt} 
\makeatletter
\newcommand\BackgroundPicturea[3]{
	\setlength{\unitlength}{1pt}
	\put(0,\strip@pt\paperheight){
		\parbox[t]{\paperwidth}{
			\vspace{#2}\hspace{#3}
			\mbox{\includegraphics[scale=0.5]{#1}}
}}}
\newcommand\BackgroundPictureb[3]{
	\setlength{\unitlength}{1pt}
	\put(0,\strip@pt\paperheight){
		\parbox[t]{\paperwidth}{
			\vspace{#2}\hspace{#3}
			\mbox{\includegraphics[scale=0.3]{#1}}
}}}
\makeatother

\usepackage{setspace}
\addto\captionsenglish{
	\renewcommand{\contentsname}%
	{Table of Contents}
	\setcounter{tocdepth}{3}
	\setcounter{secnumdepth}{3}
	}

\usepackage{xcolor}
\usepackage{listings}
\usepackage{xcolor}

\definecolor{codegreen}{rgb}{0,0.6,0}
\definecolor{codegray}{rgb}{0.5,0.5,0.5}
\definecolor{codepurple}{rgb}{0.58,0,0.82}
\definecolor{backcolour}{rgb}{0.95,0.95,0.92}

\lstdefinestyle{mystyle}{
  backgroundcolor=\color{backcolour}, commentstyle=\color{codegreen},
  keywordstyle=\color{magenta},
  numberstyle=\tiny\color{codegray},
  stringstyle=\color{codepurple},
  basicstyle=\ttfamily\footnotesize,
  breakatwhitespace=false,         
  breaklines=true,                 
  captionpos=b,                    
  keepspaces=true,                 
  numbers=left,                    
  numbersep=5pt,                  
  showspaces=false,                
  showstringspaces=false,
  showtabs=false,                  
  tabsize=2
}

\usepackage{amsfonts}

\newcommand{\sortitem}[1]{%
  \DTLnewrow{list}
  \DTLnewdbentry{list}{description}{#1}
}
\newenvironment{sortedlist}{%
  \DTLifdbexists{list}{\DTLcleardb{list}}{\DTLnewdb{list}}
}{%
  \DTLsort{description}{list}
  \begin{itemize}%
    \DTLforeach*{list}{\theDesc=description}{%
      \item[] \theDesc}
  \end{itemize}%
}

\usepackage{algorithm}
\usepackage{algpseudocode}

\usepackage{pgfgantt}

\usepackage{makecell} 

\usepackage{lscape}

\usepackage{tikz}
\usetikzlibrary{quantikz2}
\def\myvdots{\ \vdots\ }
\usepackage{adjustbox}

\def\mya{ans{\"a}tze }
\def\aa{{\'a}}
\def\ee{{\'e}}
\def\ii{{\'i}}
\def\oo{{\'o}}

\usepackage{float}

\usepackage[normalem]{ulem}
\useunder{\uline}{\ul}{}

\usepackage{blochsphere}
\usepackage[utf8]{inputenc}

%% file: Glossary.tex
\makeglossaries

\newglossaryentry{latex}
{
    name=Latex,
    text=latex,
    description={Is a markup language specially suited 
    for scientific documents, this term is printed in conclusion }
}
\newglossaryentry{raster}
{
    name=Raster,
    text=raster,
    description={ images are compiled using pixels, or tiny dots, containing unique color and tonal information that come together to create the image }
}
\newglossaryentry{gradient descent}
{
    name=Gradient descent,
    text=gradient descent,
    description={Is a naive optimization method which consists of steepest descent down the gradient of the given cost function}
}

\newglossaryentry{Gauss-Newton}
{
    name=Gauss-Newton,
    description={Is a Newton-like method for solving a non-linear least squares problem, in which the Hessian $H$ is approximated by $H \approx J^T WJ$, where $J$ is the design matrix and $W$ is the weights. The normal equations are the resulting prediction equations given as \\ $(J^TWJ) \delta x = -(JW \Delta z)$}
}

\newglossaryentry{Conjugate Gradient}
{
    name=Conjugate Gradient,
    text=conjugate gradient,
    description={Is an accelerated first order iterative process for solving positive definite linear systems or minimizing a non linear cost function.}
}

\newglossaryentry{Jacobian}
{
    name=Jacobian,
    description={Matrix of partial differentials of the cost function $J = \frac{df}{dx}$}
}

\newglossaryentry{Hessian}
{
    name=Hessian,
    description={Matrix of second partial differentials of the cost function $H = \frac{d^2f}{dx^2}$}
}
\newglossaryentry{Gradient}
{
    name=Gradient,
    description={First differential $g = \frac{df}{dx}$}
}
\newglossaryentry{epipolar plane}
{
    name=Epipolar Plane,
    description={The plane containing the intersection line joining the camera centres with the image plane.}
}
\newglossaryentry{linear least squares}
{ 
    name =Linear Least Squares,
    description ={Least squares approximation of linear functions to data, by minimizing residuals.
     $E_{LS} = \sum_i{||\hat{x'_i}- \tilde{x'_i}||}$}
}
\newglossaryentry{RANdom Sample Consensus (RANSAC)}
{
    name = Random Sample Consensus (RANSAC),
    description = {An iterative method to estimate parameters of a mathematical model from a set of observed data which contains outliers.}
}

%% file: 0_Abstract.tex
The field of Quantum Computation is plagued by issues that limit the implementation and development of quantum systems and quantum algorithms. Issues which force the development of Hybrid Quantum-Classical algorithms, such as the quantum DisCoCat implementation for Natural Language Processing. These require a high processing cost and are susceptible to errors due to Out of Vocabulary words. 

In this work, we develop a framework to implement Few Shot Learning for Quantum Natural Language Processing, by modifying the encoding \mya and dividing it into two parts, the first one leveraging the vast corpus of classical training already available, and the second variationally training on the task. This framework is then put to the test to explore its behaviour and its power in extracting as much useful work from each call to a quantum system as possible.

%% file: 1_Introduction.tex
\chapter{Introduction} \label{Chap1}
\pagenumbering{arabic}

When computers were first imagined it seemed that their practical implementation would take decades, centuries even. The first ones filled rooms and were noisy, clunky machines that could be easily outperformed by today's wristwatches. And yet, it would be hard to think of a world that dispenses with transistors and logic gates.

Quantum computers have, as of now, followed a similar path. Many attribute the notion of a computer using quantum phenomena to first come about in Richard Feynman's 1981 talk about simulating quantum phenomena. Feynman speculated about a new type of universal computer better suited to simulate quantum phenomena, prophetically describing what we would later come to know as quantum computers \cite{hey_feynman_2002}:

\begin{quote}
    "Can you do it with a new kind of computer - a quantum computer? [...] as far as I can tell, you can simulate this with a quantum system with quantum computer elements. It's not a Turing machine, but a machine of a different kind."
\end{quote} 

John Preskill described in a 2018 paper what is known as the Noisy Intermediate Scale Quantum era (NISQ). A broad and mostly qualitative term which connotes a stage in quantum computing where we are past proof-of-concept devices, but on which fidelity and scalability issues prevent us from dreaming and creating big.

Preskill defined the \textit{intermediate} adjective to refer to a stage where quantum computers were made of only 50 to 100 qubits. In December 2023, IBM unveiled their Condor computer, a 1000+ qubit computer that shows the main limitation today is not our ability to design and implement theoretically perfect quantum machines, but the \textit{noisiness} Preskill made central to the NISQ era. 

Noise is today's quantum killer. Even if there were a large number of qubits available to us, the need to perform error correction and limit decoherence renders the construction and implementation of a general-purpose quantum computer or algorithm fairly limited.

To overcome this, a few tricks have been developed. One of them is to construct Quantum Parameterised Circuits (QPC) and use Hybrid Quantum-Classical (HQC) algorithms, which involve a system of classical and quantum computers working together to solve a problem. These tools allow for the exploitation of quantum computers in the NISQ era and have ushered in the development of new techniques such as Quantum Machine Learning (QML).

Quantum Natural Language Processing (QNLP) is a sub-field of this last discipline that focuses on translating and implementing classical Natural Language Processing (NLP) techniques to quantum circuits.

Even though QPCs and HQC algorithms help to overcome many of the issues brought on by noisiness, there is still room for improvement if we want to squeeze out every drop of performance from a quantum processor. Few-shot learning (FSL) is a technique in classical Machine Learning that seeks to extract a reasonably useful training cycle in an environment where training capabilities are somewhat limited. The constraints on which FSL operate make it an ideal candidate to implement in QML to be able to obtain more out of each call to a quantum computer.

Quantum computers are a reality, and they might be big and clunky machines that require delicate touch and clever manoeuvring to show their true potential. But as many computer scientists did more than half a century ago, we strive to construct a world where a future without qubits becomes practically impossible to imagine.

%% file: 2_Scheme.tex
\chapter{An Outline} \label{Scheme}

The contribution of this project to the field of QNLP is the development of the framework of Few Shot Learning to be used in running DisCoCat circuits and the testing to show its usefulness. This framework allows for the reduction in training resources at a marginal cost in accuracy that decreases with increasingly bigger circuits.

To set this up, we will first set the minimum theoretical background in chapter \ref{chap:Theory}, where the fields of Quantum Machine Learning, and Quantum Natural Language Processing will be covered, along with a description of DisCoCat and Few Shot Learning.

Chapter \ref{chap:Problem} will describe the specific problem we are trying to solve, along with the motivations for choosing and solving that specific problem.

Following an explanation of the framework in chapter \ref{chap:framework}, chapter \ref{chap:experiment} will detail the experimental design to test how this framework compares to traditional methods as well as the results of this testing. 

Finally, in chapter \ref{chap:conclusion} we will summarise what the experiments tell us about the framework, as well as its limitations, and what possible experimental avenues could be further explored.

Further information can be found in the appendices, including more information on bias in ML, the notion of quantum supremacy, and other interesting areas in QML which could benefit from FSL. All the code is linked to in the appendix as well.

%% file: 3_Theory.tex
\chapter{Theory} \label{chap:Theory}

In this chapter, a brief outline of QPCs, HPC algorithms, and QML will be delineated. Then a description of QNLP and FSL in the classical and quantum implementations will be given.

\section{Computation in the NISQ era}\label{sec:QPC,HPC}

The NISQ era sets certain limitations on what we can practically do with quantum computers (for those outside the field see \cite{lau_nisq_2022}). Ideally, we would have an all-purpose computer consisting of $N$ qubits on which we could perform arbitrarily many n-qubit unitary operations (where $n \leq N$). However, a well-known result indicates that the minimum number of gates needed to perform an arbitrary N-qubit unitary grows to the order of $4^N$ \cite{barenco_elementary_1995}.

Each additional qubit comes at a cost in increased noise due to the imperfections in the device \cite{krantz_quantum_2019} and the qubits' interaction with the environment \cite{georgopoulos_modeling_2021}. Each additional single and two-qubit gate also brings with itself additional noise \cite{di_bartolomeo_noisy_2023}. All of these make error-correcting a more daunting task that reduces the benefit of having many more qubits \cite{knill_theory_2000}.

With quantum advantage being limited to shallow circuits \cite{bravyi_quantum_2018}, the problem in the NISQ era reduces to developing useful algorithms with a reduced amount of qubits, as few calls to a quantum computer as possible, and as few unitaries as possible.

\subsection{Variational and HQC Algorithms}

One of the most common ways to extract work from quantum computers is through HQC algorithms. These are a class of algorithms that separate each problem into sections and then assign them to a classical or quantum system so that each system works on the type of problem they have an advantage \cite{mcclean_theory_2016}.

The most commonly used HQC algorithms are variational algorithms, where a problem is encoded into a cost function best suited for evaluation in a quantum computer \cite{wecker_progress_2015}. This cost function is encoded through a PQC, where the unitary transform comprises of rotation and controlled rotation gates whose angles are free to change, hence the variation. The quantum circuit is evaluated and then a classical optimiser is used to calculate the adjustment to the rotation parameters to either minimise or maximise the cost function \cite{cerezo_variational_2021}. 

The go-to example for variational algorithms is the Variational Quantum Eigensolver (VQE), first described in 
\cite{peruzzo_variational_2014}. It is one of the widely used variational algorithms. Its power lies in encoding the problem into a Hamiltonian whose solution is the minimum eigenvalue, the ground state \cite{tilly_variational_2022}. The classical optimisation algorithms find the quantum ground state which is the solution to the problem. It has been widely used to find the ground state of molecules whose analytical and classical numerical solutions are too complicated or costly to otherwise obtain \cite{kandala_hardware-efficient_2017}.

Variational algorithms have been proven to be resilient to noise \cite{sharma_noise_2020,ding_evaluating_2022,fontana_evaluating_2021} and to have incredible versatility for the classes of problems they can be applied to \cite{cerezo_variational_2021}.

Other variational algorithms include the Quantum Approximate Optimisation Algorithm (QAOA) \cite{farhi_quantum_2014}, Quantum Auto Encoder (QAE) \cite{romero_quantum_2017}, and Variational Quantum Error Mitigation (VAQEM) \cite{ravi_vaqem_2022}.

Given all these benefits, it is reasonable to ask, what's the catch? The design and implementation of HQCs and variational algorithms come with an interesting problem:

\subsection{What would be the best way to construct a parameterised circuit?}\label{subsec:ansatz}
\subsubsection{Ans{\"a}tze}

If we consider a general problem there is almost complete flexibility in the variational circuit we construct for it, both in the number of qubits and the depth (the number of layers) of the circuit.

An ansatz is a general structure of ordered single and multiple qubit gates that can be extended to any arbitrary number of qubits and repeated in layers to an arbitrary depth (circuit \ref{circ:an}).

It can be expressed through a set of unparametrised gates ${V_l}$, such as the set of Pauli X, Y, and Z gates, with a set of unitaries parameterised by a set of angles $\boldsymbol{\theta}={\theta_1,\theta_2,...,\theta_n}$ and traceless rotational generators $V_l$ \cite{leone_practical_2024}.

\begin{align} \label{eq:genansatz}
    U(\mathbf{\theta})=\prod_l e^{-i\theta_l H_l} V_l
\end{align}

\input{circuits/genan}

The most straightforward parametrisation is that of a single qubit. The Euler parameterization ansatz for a single qubit (for a general SU(N) group see \cite{tilma_generalized_2002}) consists of applying to a single qubit \cite{tilma_applications_2003} three consecutive rotation gates (circuit \ref{circ:ep}), Rz, Ry, Rx: $|\Psi \rangle = U(\theta_z,\theta_y,\theta_x)|0\rangle = \textbf{Rz}(\theta_z)\textbf{Ry}(\theta_y)\textbf{Rx}(\theta_x)|0 \rangle$. This ansatz can produce any single-qubit quantum state. Visually, it can be pictured as a $\theta$ radians rotation of a qubit state on the Bloch Sphere around an arbitrary axis $\hat{n} =x \hat{\textbf{i}}+y\hat{ \textbf{j}}+z\hat{ \textbf{k}}$ (figure \ref{fig:eulerparam}). Since the analytical calculations to obtain the set of rotation angles for any state are fairly straightforward and computationally simple, this ansatz is enough for simple toy tasks. 

\input{circuits/euler_parametrization}

\input{circuits/ep}

The Euler parameterization is also non-scalable. That is, this ansatz is only defined for a single qubit and a single layer (as it is an SU(2) parametrisation). For bigger circuit sizes, we would need gates that can produce an entangled state, so gates that involve two or more qubits \cite{nielsen_quantum_2012}.

The most simple ansatz for two or more qubit circuits is called the Instantaneous Quantum Polynomial (IQP) ansatz. This ansatz, as the Euler Parameterisation, is used mostly for toy examples and proof-of-concept experiments, and is, in reality, a family of \mya diagonal in the X basis and which permits the sampling of the state distribution in polynomial time \cite{bremner_average-case_2016}. A single ansatz layer (circuit \ref{circ:IQP}) is defined as a row of Hadamard gates and then a row of Controlled-Z rotations applied to nearest neighbour qubits:
\begin{align} \label{IQP}
    | \Psi \rangle = \textbf{CZ}(N-1,N,\theta_N{-1})...\textbf{CZ}(1,2,\theta_2) \textbf{CZ}(0,1,\theta_1) H^{\otimes N}|0 \rangle,
\end{align}
where $\textbf{CZ}(i,j,\theta_i)$ is a Z-rotation of $\theta_i$ on qubit j controlled by i.

\input{circuits/iqp}

The previously mentioned \mya serve mostly as an illustration of the concept and for usage in simple problems. Other ans{\"a}tze are more generally used and are usually designed for specific use cases.

The Hardware Efficient Ansatz (HEA) \cite{kandala_hardware-efficient_2017} is a class of \mya where the specific implementation considers the specific hardware requirements of the quantum computer it intends to run on. The HEA considers the qubits to be arranged in a ring-like topology, where each qubit $i$ is connected to its nearest neighbour $i+1$ and the last and first qubits are connected (figure \ref{circ:hea}). It is intended to be an ansatz that allows efficient connectivity while also being depth-frugal \cite{park_hardware-efficient_2024}.
\input{circuits/efficient}

Other \mya such as Sim15 Ansatz \cite{sim_expressibility_2019} and Strongly Entangling Ansatz \cite{schetakis_binary_2021}, have their merits and their implementations whenever the problem might call for.

\subsubsection{Expressibility and Entanglement Capability}

But the reality is that the creation of an ansatz has few real limitations. As long as the gates can be implemented (or simulated if we don't have access to quantum hardware) there is no practical reason why an unpractical ansatz couldn't be thought of and implemented.

When searching for an ansatz for a variational algorithm, the considerations taken into account are usually hardware-related. As mentioned in the HEA, this usually means how the qubits are topologically connected as well as the specific qubit engineering, some indicators are needed to guide the search for a particular ansatz that transcends necessarily practical concerns. These indicators might be useful when comparing similar \mya or in the early stages of algorithm development when the performance is most important. In this context two measures were developed as a guide in ansatz selection: expressibility and entangling capability \cite{sim_expressibility_2019}.

Expressibility is the measure of how much of the Hilbert space where the quantum state lives is reachable. At the beginning of section \ref{sec:QPC,HPC} we mentioned that for a general $N$ qubit unitary, we would need at minimum around $4^N$ gates. It is important to remember that the goal of any variational algorithm is to find the parameters that make the ansatz behave as a unitary transform $U$ that solves the problem. But the amount of gates used for any ansatz grows polynomially, not exponentially, so even though an ansatz might approach the ideal $U$, it is not a given that it will reach it close enough to be within a margin of error defined.

Expressibility is defined formally as \cite{sim_expressibility_2019} \begin{quote} "a circuit's ability to generate (pure) states that are well representative of the Hilbert space",
\end{quote} which for a single qubit is essentially\begin{quote}"a circuit's ability to explore the Bloch sphere".
\end{quote}

In that same paper it is estimated using the following formula:
\begin{align}
    \text{Expr}=D_{KL}(\hat{P}_{PQC}(F;\boldsymbol{\theta})\parallel P_{Haar}(F)),
\end{align}
where the Kullback-Leibler (KL) \cite{belov_distributions_2011} divergence is calculated between randomly sampling states from the PQC and calculating the estimated distribution of fidelities ($\hat{P}_{PQC}(F;\boldsymbol{\theta})$), and the analytical form of the probabilities distribution ($P_{Haar}(F)$). Using this scheme, a lower KL divergence indicates a more expressible circuit.

Entanglement capability is another useful metric by which to judge \mya. Many algorithms necessitate the preparation of the initial state to be in a certain superposition of pure states, so having an ansatz which can approximately provide this independently of the problem it encounters becomes a necessity for these algorithms in the NISQ era.

Even though other entanglement measures exist and might be more widely used, such as von Neumann's entropy of entanglement \cite{boes_von_2019} or the Schmidt number \cite{terhal_schmidt_2000}, in the paper it is formally defined using the Meyer-Wallach measure \cite{meyer_global_2002} because of its "scalability and ease of computation". Since in this work, the entanglement capability is not used, we will just give the mathematical definition as a starting point for those interested:

For a linear mapping $\iota_j(b)$ on the computational basis: $\iota_j(b)|b_1\cdots b_N\rangle=\delta_{b,b_j}|b_1\cdots\hat{b}_j\cdots b_N\rangle$, where $\hat{b}_j$ is absent. the MW entanglement measure Q is:
\begin{align}
    Q(|\Psi\rangle)=\frac{4}{N}\sum_{j=1}^n D(\iota_j(0)|\Psi\rangle,\iota_j(1)|\Psi\rangle),
\end{align}
where $D(|u\rangle,|v\rangle)=\frac{1}{2}\sum{i,j}|u_i v_j-u_j v_i|^2$.

It is not straightforward how a particular ansatz will perform in any of these measures, and the expressibility and entanglement capability may change depending on the number of layers used and the width of the circuit. The rate at which it might change between layers is not necessarily linear or consistent across ans{\"a}tze. When selecting a particular ansatz expressibility and entanglement capability should not be taken just as single numbers that can tell the whole story on the architecture, but as a guide that should be taken into account along with all of the requirements of resource usage and any other that might appear. To end with a brief qualitative example, if we would like to use quantum hardware with a ring-like topology, we would look for a version of the HEA that has the highest expressibility within the first two layers, even though others might exist that performs asymptotically better when more layers are added.

\section{Making Qubits Learn}
Quantum algorithms have come a long way since the early days of the field, where their only purpose was to show that Feynman's quantum machine could indeed outperform their classical counterparts.

The first algorithms were of the toy variety, simple problems that often required one or two-qubit registers and very specific circumstances. Deutsch's \cite{deutsch_quantum_1985} (and later Deutsch-Jozsa's \cite{jozsa_role_2003}) algorithm where the task is to find if a function $f(x)$ is constant or balanced (i.e. outputs always a 1 or a 0 or each half the time) showed exponential speedup compared to its best classical counterpart. Grover's search algorithm \cite{jozsa_searching_1999} needed $\sqrt{N}$ calls when compared to the classical version.

The Quantum Fourier Transform (QFT) \cite{weinstein_implementation_2001} and Quantum Phase Estimation (QPE) \cite{jiang_survey_2021} had to arrive for the field to see the advantage quantum computers had on practical problems. Most notably, Shor's factorisation algorithm leveraged quantum computers to break prime factorization and thus many modern encryption schemes such as RSH Cryptography \cite{wong_shors_2024}.

It is worth mentioning that quantum teleportation \cite{pirandola_advances_2015} and dense coding \cite{guo_advances_2019} do provide useful results, but it is the opinion of the authors of this work that they show that what a quantum computer can do is not necessarily linked to or has to be compared to a classical computer, rather they are proof that the future of the field lies far off the classical path. Ironically, this makes them somewhat unsuited for discussions of the NISQ-era algorithms.

\subsection{Quantum Machine Learning}

Quantum Machine Learning is a subfield that deals with the advantages quantum computers bring to traditional ML. This field rests on the notion that if quantum computers can produce statistical patterns that aren't easily or efficiently reachable for a classical processor, then a quantum computer might also be able to recognise statistical patterns that a classical computer might not \cite{biamonte_quantum_2017}.

What is efficiently computable for a quantum system is not determined by a simple discussion. Current arguments for quantum supremacy rely not necessarily on a strong mathematical proof as in the classical relative of the field, but in a discussion of the state of the art and the comparison with their classical counterpart. More on quantum supremacy can be seen in appendix \ref{app:QuantumSip}.

As of the time of writing, most of the literature in QML relies on variational algorithms and PQCs, where, for example, a variational circuit outputs a state in a space where the basis states are mapped to labels in a classification task. The cost function then relates to how well the circuit classifies a dataset, and the data is encoded in the parameters of the circuit. Then the classical teammate of the HQC algorithm would be tasked with modifying the parameters until the encoding is good enough for a dataset.

Since this work is mainly on Quantum Natural Language Processing (QNLP), to end this section we will just give a couple of common QML architectures for the curious, and delve deeper into QNLP in the next section.

Quantum Neural Networks (QNN) are the quantum counterpart of the classical notion of neural networks. The term QNN has referred to a lot of architectures through the years, where the common ground is the adoption of a particular characteristic of neural networks. For example, the notion of feed-forward neural networks with non-linear activation functions lends itself particularly well to photonic implementations of quantum computers \cite{killoran_continuous-variable_2019,steinbrecher_quantum_2018}. The vast majority of references use the term QNN to refer to variational algorithms and use them in classification tasks \cite{farhi_classification_2018}, where the driving similarity is the modularity and interconnectedness of classical neural networks. For a deeper dive into the theory of QNN, we refer to \cite{schuld_quest_2014}. The use cases of QNN can be thought of as similar to those of classical neural networks. Some examples are: classifying handwritten digits \cite{zhou_recognition_1999}, health applications in oncology \cite{li_model_2014} and cardiology \cite{jie_zhou_automatic_2003}, and time series forecasting \cite{azevedo_time_2007}.

Quantum Generative Adversarial Networks (qGAN) follow the same principle as classical GANs. The task is twofold, with a discriminator and a generator, the discriminator should create a probability distribution with some input that should closely resemble a target distribution. The discriminator must then tell if any given distribution is the target one or the one created by the generator \cite{goodfellow_generative_2014}. The key to qGANs is the creation of an additional quantum circuit used to calculate the gradients needed to optimise the generator circuit's parameters \cite{dallaire-demers_quantum_2018}. qGANs have applications in finance \cite{zoufal_quantum_2019}, chemistry \cite{kao_exploring_2023}, options pricing \cite{fuchs_hybrid_2023}, among others.

Motivated by the benefits of convolutional neural networks, quantum convolutional neural networks (QCNN) seek to emulate the structural characteristics in the quantum realm. In a QCNN, the convolution is constructed through \cite{cong_quantum_2019} "a single quasilocal unitary ($U_l$) in a translationally invariant manner for finite depth", and pooling and nonlinearity is done through measuring some qubits and using the results to adjust the parameters of the subsequent layers. QCNN can be used, as traditional convolutional neural networks, in image classification, as well as in other broader uses such as fluid dynamics problems \cite{umeano_what_2024}.

\subsection{Quantum Natural Language Processing}

\subsubsection{Classical Natural Language Processing}
Many things have come from Douglas Adams' famous five-book trilogy The Hitchhiker's Guide to the Galaxy. One of the earliest and now only known to a niche group of enthusiasts is the computer game of the same name. This game came out in 1984 and was an interactive fiction video game. This genre of games comes from an era where computers were operated only through the command line and relied on the player reading the description of the scene they were in, and describing in the command line what their actions should be. Normal inputs would be of the type "Go north", "Open box", or "Look around", and if the player dared to stray too far from the interpreter's capabilities they would be rewarded with a message of the type "I do not understand that command".

In a sense, these early games embodied what the field of Natural Language Processing (NLP) strives to achieve, a system that can parse and operate on a user's input as if they were two people speaking.

The popular imagination of the field of NLP has been saturated with recent developments specifically in Large Language Models (LLM) brought upon by transformers \cite{chang_survey_2024}. LLMs now permeate modern life, from companies adapting OpenAi's ChatGPT to serve as assistants and the face of their customer service departments to Meta unilaterally unleashing their Lama model on Whatsapp's users as just another member of their group chats.

However, NLP covers a more diverse area than the creation of generative pre-trained models (GPT). NLP covers a field as diverse as meaning classification \cite{tsvetkov_evaluation_2015}, sentiment analysis \cite{schnabel_evaluation_2015}, speech comprehension, paraphrasing \cite{baumel_sentence_2016}, pronoun resolution \cite{wazni_quantum_2022}, and anything that has to do with how humans communicate \cite{chowdhary_natural_2020}.

For the specific uses of this work, we want to deal with how meaning as a whole arises from the meaning of the specific parts of speech and how they are composed in the structure that binds them together. That is, we are interested in a compositional model of language, where we care not only about the meaning behind the words but also about their structure, for more on the general field of NLP see \cite{klontzas_natural_2023,nadkarni_natural_2011}. What follows is a brief but sufficient explanation of word embeddings, compositionality, and the categorical compositional distribution semantic (DisCoCat) model of language and how we can seamlessly translate it to a quantum implementation.

\subsubsection{Word Embeddings} \label{sec:embeddings}
The first step to developing a compositional model of meaning is to find a way to assign each word a mathematical object that represents their meaning and which can be used to compare them.

A word embedding is the representation of the meaning of a word in the context of its survey space \cite{jiao_brief_2021}. To unpack this, we first consider what this representation can be. Usually, each word is assigned a vector whose vector space could be thought of as the meaning space of all the words. However, it does not need to be a vector. It can be an N-rank tensor that takes into account other factors, such as the grammatical role of a word \cite{kartsaklis_prior_2013,rahimi_tenssent_2021,rahimi_tens-embedding_2020}. Each tensor is then calculated using a survey space, or a sample of many texts where words can be compared to one another in terms of frequency and how they appear with each other \cite{bakarov_survey_2018}.

From this method of embedding calculation, based on the distributional hypothesis (a hypothesis that proposes that words with similar meanings appear in similar contexts \cite{lenci_distributional_2018}), arise some interesting results. For example, if we take the vector embedding for king, subtract the vector embedding for man, and add the vector embedding for woman, we end up with the vector embedding for queen \cite{mikolov_linguistic_2013}:
\begin{align*}
    \overrightarrow{king}-\overrightarrow{man}+\overrightarrow{woman}=\overrightarrow{queen}
\end{align*}
This suggests that from some vector set that spans the embeddings vector space, one of those corresponds to the encoding of the meaning of gender, and so it might be that other vectors in some other spanning set could encode other fundamental meanings of words, for example expanding a word vector into a weighted sum of the meaning basis as:

\begin{align}
    \overrightarrow{dog}=A*\overrightarrow{animal}+B*\overrightarrow{fur}+C*\overrightarrow{friendliness}+...
\end{align}

Word embeddings form the basis for many more complex NLP tasks, but from them, some rudimentary analysis can be performed. One can judge how alike two words are in meaning by calculating their distance using some metric (usually Euclidean distance but others can be used), those words whose meaning is similar should be closer to each other \cite{che_traversal-free_2017}. The cosine distance is another technique used \cite{mikolov_linguistic_2013}, it is just the dot product of two embeddings and provides similar information.

The way to obtain these embeddings can vary in anything from the sources themselves, Stanford's GloVe embeddings \cite{pennington_glove_2014}, for example, are a set of embeddings calculated from three different sources: Wikipedia, Common Crawl, and Twitter, to the different techniques used to extract the embeddings. Co-occurrence is usually the main driving principle for any algorithm, but modifications in the unsupervised models do exist. A recent technique called BERT (Bidirectional Encoder Representations from Transformers) \cite{devlin_bert_2019}, treats each word string as a single entity. Previous models usually read each word directionally (either left to right or right to left). 

That these models rely on data created and consumed by humans makes them quite susceptible to bias, a brief discussion can be seen in appendix \ref{app:bias}.

\subsubsection{Compositionality and Bag of Words}
The language we use does not rely on individual words to tell whole ideas. Language is based on compositionality \cite{moore_meaning_1993,riemer_routledge_2016}, words are used together to form complex text structures that alter the individual meaning of each of their components. So to construct a model capable of doing more interesting tasks, a compositional model is needed atop the word embeddings previously described.

The Bag of Words (BoW) model of language \cite{qader_overview_2019,zhang_understanding_2010} comes from distributionality. One of the key features of language modelling since the middle of the 20th century, Distributionality is the principle that what matters most to identify the meaning of a piece of text or a sentence is the existence and multiplicity of its atomical components \cite{harris_distributional_1954}. It is called Bag of Words due to its visual analogue of throwing all the words into a bag as if they were Scrabble tiles. The order and their relation are lost.

As a model, BoW is simple and not computationally intensive. It is also versatile, being adapted to computer vision \cite{qader_overview_2019}, as there are also models that can transform an image into an embedding that encodes its features. It suits particularly well that field as computer vision is resource intensive.

However, what makes it simple is also one of its greatest drawbacks. The syntactic structure of text also forms part of its meaning \cite{ives_notes_1964}. For example, even though "Boys like dogs" and "Dogs like boys" have the same word tokens, they mean different things. For simple tasks, this does not affect performance significantly. Nevertheless, a more complete model of language is needed once we try to expand the use cases of NLP and minimise its errors.

And thus we arrive at DisCoCat.

\subsubsection{DisCoCat}\label{sec:discocat}

For a complete theoretical description, we point to the paper on which DisCoCat was originally introduced \cite{coecke_mathematical_2010}. A brief but sufficient description of the mathematical model follows, which serves as enough foundation to understand its quantum implementation. Briefly, DisCoCat gives a method to obtain the meaning vector $\Vec{s}$ of any sentence, where for any sentence they all live in the same meaning space $\{ \Vec{s}\}_i \in S$, thus allowing the comparison of the meaning of sentences with different grammatical structures.

DisCoCat follows a categorical-theoretic approach to model language. The authors justify this choice by enlisting what a compositional model of language needs and how this is fulfilled by categories. Firstly, the structure of monoidal categories helps capture compositionality, particularly through the monoidal product. This in turn helps to leap from a qualitative space of meaning to a quantitative one. The structural morphisms of the chosen categories help shape the morphisms to construct the flow they call, from-meaning-of-words-to-meaning-of-a-sentence. Finally, this model helps reason about the grammatical structure of a sentence as an object in and of itself, thus allowing us to study more in-depth the role grammar plays in endowing a sentence with its meaning.

The from-meaning-of-words-to-meaning-of-a-sentence model gives us a recipe to use the categorical-theoretic machinery to take some vector representation of the meaning of the individual words and how they are related dramatically to each other to output the meaning of the sentence. 

They do this by first stating two categories: \textbf{FVec}, which is the category with vector spaces over the field of reals $\Re$, and the tensor product $\otimes$ as the monoidal tensor, and the category of Pregroups, \textbf{P}, a posetal category, endowed with the morphisms needed to turn it into a compact closed category. Each of them serves a specific purpose. \textbf{FVec} is in charge of modelling the meaning of the words, and \textbf{P} assigns the grammatical relationship of the words in the sentence.

Instead of having the meaning of the word be just that of its embedding vector, the authors define the meaning space of the word $\Vec{w}$, element of the vector space $W$, $\Vec{w} \in W$, as an object $(W,p)$ of \textbf{FVec}$\times$\textbf{P}, where p is the grammatical type of $\Vec{w}$. So the meaning of a string of words ordered in some particular way is a linear map $f$, $\overrightarrow{w_1\otimes w_2\otimes \cdots \otimes w_i}:=f(w_1w_2\cdots w_i)$, which relies the structural morphism: 
\begin{align}
    (W_1\otimes W_2 \otimes \cdots \otimes W_i,p_1p_2\cdots p_i)\rightarrow^{(f,\leq)}(X,x).
\end{align}
The map $f$ is obtained by substituting each $p_i$ in $p_1p_2...p_n\leq x$ with $W_i$. This map $f$ is the core of the model and is the from-meaning-of-words-to-meaning-of-a-sentence map.

So to obtain the meaning of some sentence, the recipe involves first assigning a grammatical type to each of the words in the sentence. Then defining the vector space which encodes the meaning of each word, and finally taking the tensor product of the words and applying to it the map of the syntactic reduction of the string.

To end with an abstract example, we will look at what happens when we try to find the meaning of a positive transitive sentence. This is a sentence of the form Subject-Verb-Object, e.g. I love you. We start by defining the meaning spaces of the subject and the verb as $(Sub,n)$ and $(Ob,n)$, respectively (one can see both share the same grammatical type n as they are both nouns). The meaning space of the transitive verb $T$ is $(Sub\otimes S    \otimes Ob,n^rsn^l)$. The meaning map $f$ we are looking for is the map that realises the structural morphism:
\begin{align}
    (Sub\otimes T \otimes Ob,n(n^rsn^l))\rightarrow^{(f,\leq)}(S,s),
\end{align}
which according to the syntactic reduction map is
\begin{align}
    f=\epsilon_{Sub}\otimes 1_S \otimes \epsilon_{Ob}:Sub\otimes(Sub\otimes S    \otimes Ob)\otimes Ob \rightarrow S,    
\end{align}
and to get the meaning vector we would apply $f$ to the tensor product of the embeddings of the sentences:
\begin{align}
    f(\overrightarrow{Sub}\otimes \Vec{T} \otimes \overrightarrow{Ob})=\epsilon_{Sub}\otimes 1_S \otimes \epsilon_{Ob}(\overrightarrow{Sub}\otimes \Vec{T} \otimes \overrightarrow{Ob})
\end{align}
To throw some numbers around, say we define the meaning space of a sentence to have its vector space be spanned by $\hat{e}_1=[1,0]$ and $\hat{e}_1=[0,1]$, where $\hat{e}_1$ is the meaning vector we assign to the word true, and $\hat{e}_2$ is the one we assign to false. We then would have two sentences: $\Vec{s_1}=\text{I love you}$ and $\Vec{s_1}=\text{My friend likes me}$. After defining the proper grammatical pregroups and applying the corresponding meaning map, we would obtain the meaning vector $m_1=f(\overrightarrow{s_1})=[0.90,0.28]$ if you are my girlfriend and $m_2=f(\overrightarrow{s_2})=[0.35,0.8]$ if I might have missed my friend's birthday party. This will indicate that, according to our definitions of the spanning vectors of $S$, sentence 1 is truer than sentence 2. We can not only assert whether each sentence is true or not, but we can also compare them to one another even though they have different words and syntactic structures.

This is the power and advantage of \textit{classical} DisCoCat.

\subsubsection{Schr{\"o}dinger's DisCoCat}
This section builds on all the theories we have discussed so far.

We have talked about vectors, tensor product, and, in a sense, the flow of information across a sentence. In \cite{coecke_mathematical_2010} the authors even make heavy use of diagrammatic language that if one were to squint and maybe turn their head on their side it might look like the diagram of a quantum circuit. Indeed DisCoCat is a compositional model of meaning that, under the hood, relies computationally on tensor networks. So it is natural to ask if there might be some way to do DisCoCat on a quantum computer.

The process of turning a DisCoCat diagram into a quantum circuit is straightforward and outlined in \cite{lorenz_qnlp_2023,meichanetzidis_quantum_2021}. In a short, but mathematical description, functors are used to translate between the categories of Pregroup grammar \textbf{P} and finite Hilbert spaces \textbf{fHilb}, $\mathbf{F}:\mathbf{P}\rightarrow\mathbf{fHilb}$. Then we need to choose a concrete way in which the circuit will take shape through an ansatz. This in turn has its considerations mentioned in section \ref{subsec:ansatz}. The mapping into the category \textbf{fHilb} means that each type \textbf{P} will have its own dimensionality. So, for example, a noun of type N will correspond to a Hilbert space of dimension 4, while a sentence of type S will correspond to a Hilbert space of dimension 2. After the ansatz is chosen, we will need to construct the circuit according to the rules of DisCoCat, which is, essentially, turning the wires in the classical diagrams into wires in the quantum circuit. After all the mappings are finished and we have a fully connected circuit, we just run it and measure it to obtain the state representing the meaning of the sentence.

Some more post-processing is required. After taking a certain number of shots, some of the measurements that do not fit certain requirements need to be discarded. Currently, the state of the art for quantum DisCoCat (and its classical implementation) relies on HQC and variational algorithms, so in reality, we are training the parameters of a PQC to learn the meaning space of the sentence and to correctly output the state given a variety of words and their order.

This happens in classical implementations of DisCoCat as well. According to DisCoCat, it is reasonable to define before any processing is done both the meaning vectors of the words and the meaning space of the sentence. So both the word encodings are learned and the sentence meaning space is defined through the training data.

\subsubsection{The Bad News}

Given the direct functorial translation between classical DisCoCat and an abstract implementation into a PQC, the difficulty lies not in how to perform NLP, but how to perform it efficiently. As mentioned before, current methods involve training a circuit with a given ansatz to perform a given task, usually with basic ans{\"a}tze. For example in \cite{wazni_towards_2023} the authors trained a circuit using the IQP ansatz to perform pronoun resolution.

There is, as of now, no direct DisCoCat implementation on a quantum circuit that doesn't rely on training it. This comes at a great cost in both the running of the circuits themselves and the classical optimisation needed to train them. After training, these circuits have also other shortfalls. There is no reliable way to obtain the output of a circuit with out-of-vocabulary (OOV) words, words that the engine didn't see during training.

Given the vast amount of resources needed to train a quantum computer for an NLP task, there is a real need to maximise each shot and offload all the work to a classical computer, so that a PQC does what it knows best.

\section{Few Shot Learning in QML: the why the how and the who}
\subsection{The why}
The problem of cost is not endemic only to QNLP, it is a product of the NISQ era. Running a quantum computer is expensive \cite{chauhan_quantum_2022}, simulating a quantum computer is expensive \cite{zhou_what_2020}. As HQC algorithms do, the name of the game is offloading as much work as possible to classical computers

We can treat quantum computers as a resource that we have in limited quantities, and search for classical frameworks which seek to maximise the performance of ML models in circumstances of reduced resources.

Few Shot Learning (FSL) is a technique in ML that helps machines label objects using as few resources as possible by exploiting the known structure of some previous categories \cite{li_fei-fei_one-shot_2006}.

FSL is an umbrella term now that encompasses a host of techniques that in some way or another, manipulate the datasets, the models themselves, or the training methods to maximise the performance of the models in circumstances of reduced resources. The literature is vast and we will point to some sources for a more in-depth look \cite{parnami_learning_2022}. For this work, we will limit ourselves to explaining the relevant aspects of FSL that could provide some benefit to QML.

\subsection{The how}
Literature and techniques in FSL can be sorted according to the types of problems they help to overcome and according to how previous knowledge is used to solve them \cite{wang_generalizing_2021, liu_embedding_2022}. From them, here is a survey of the techniques we consider would specifically benefit QML and NISQ-era computing.

\textbf{Transfer learning} is a technique that promises to leverage the vast amount of information obtained through classical ML. This technique tries to leverage some amount of information that is abundant in some domain, to perform tasks in another domain where information might be scarce \cite{pan_survey_2010,niu_decade_2020,zhuang_comprehensive_2021}. This technique becomes particularly useful if we want to try to find a way in which we could use classical embeddings to guide the training of the variational parameters. Classical embeddings are vast and easy to compute, so if we could find a way to leverage them, then the number of times a given circuit would need to be computed would decrease.

\textbf{Imbalanced learning} is applied in cases where the dataset is askew, for example, if we are trying to have a model classify between five different classes and most of the dataset corresponds to only two classes \cite{herbelot_high-risk_2017,haibo_he_learning_2009}. This applies to NLP meaning classification tasks (and by extension their quantum counterpart) where not only the dataset might be askew, but also in the cases where some words might appear heavily in some classes even though they would apply to more than one.

\textbf{Meta learning} can prove to have some applications in QNLP. Its core idea is the construction of a meta-learner, a model whose job is to identify common structures among tasks that the learners then use to enhance their training \cite{goos_learning_2001,hospedales_meta-learning_2021}. This might be useful in QNLP since embeddings are constructed through the definition of a structure. If a task could share a similar Hilbert space on which the meaning vector space lives, then a meta-learner would be a useful tool.


The implementation of any of these methods can also vary, and it is useful to classify them according to which part of the training process they influence \cite{wang_generalizing_2021}: the data, the model, or the algorithm. The first one is the most straightforward. Given a dataset, the FSL methods that focus on that stage try to augment it to get more samples if the samples are limited (for an image model we can manipulate each sample to add multiple instances of it to the data set) or to clean the existing samples. In the algorithm branch, we modify the learning algorithm to nudge it in the direction we already know the solution should be. Finally, in the model part, we artificially restrict the search space of the model to forbid it from wasting resources exploring parts of it from previous knowledge we know contain no solution.

Each of the three might have a specific role in QNLP and QML. The particular emphasis on embedding and multitasking learning the model approach has, for example, can make it particularly suited for QNLP. The data approach is also relevant due to its applications in transfer learning. Finally, the algorithm becomes relevant when considering the meta-characteristics of QNLP and QML, especially the barren plateau problem \cite{friedrich_avoiding_2022,holmes_connecting_2022,wang_noise-induced_2021,cerezo_cost_2021,patti_entanglement_2021,mcclean_barren_2018}. For this work, it suffices to say it helps us refine the learning processes of our models.

\subsection{The who}\label{sec:who}
One of the most promising results for FSL in QML is the 2022 paper by Liu et al \cite{liu_embedding_2022}. As in section \ref{sec:discocat}, we will touch upon the most important aspects of the paper and recommend its full reading.

In their paper, they propose a framework to learn embeddings based on the paradigms of classical FSL. They proposed the construction of a parametrized circuit divided into two segments (figure \ref{circ:pqe}): first, an encoding parametrized layer whose parameters are obtained through some (not necessarily linear) mapping of classical available data to rotation angles, this layer is called the Pre Quantum Embedding (PQE) layer, and in the paper they trained a neural network to map an image of a character to the gates' rotation angles; then layers of what could be thought of as variational circuits whose parameters are the ones that are updated through the model's learning cycles.

\input{circuits/pqe}

Using this method they encoded two different images of characters and used the output of a circuit, a real number between 1 and 0, to gauge whether the model thought the images represented the same character. Their paper showed promising results: they found that FSL can generalise to unseen classes, allow for a more in-depth exploration of the solutions space, and outperform classical cosine distance. This first experiment shows that FSL can work in HQC and variational algorithms.

As of the moment of writing, FSL for NLP tasks on a quantum circuit has yet to be implemented.

Finally, it is necessary to mention a recent result which shows that a model with $T$ trainable gates and $N$ training data has a generalisation error that scales at worse to the order of $\sqrt{T/N}$, and which in some cases can be improved to $\sqrt{K/N}$ where $K\ll N$ \cite{caro_generalization_2022}.

%% file: circuits/genan.tex
\begin{figure}
    \centering
    \caption{Structure and usage of a general ansatz}
    \label{circ:an}
    \begin{quantikz}
        & \gate[6]{U(\mathbf{\theta_i})}\gategroup[6,steps=1,style={dashed,rounded corners,fill=blue!20, inner xsep=2pt},background,label style={label position=below,anchor=north,yshift=-0.2cm}]{Base layer}&\gate[6]{U(\mathbf{\theta_j})}\gategroup[6,steps=4,style={dashed,rounded corners,fill=yellow!20, inner xsep=2pt},background,label style={label position=below,anchor=north,yshift=-0.2cm}]{Base layer repeated m times}&\gate[6]{U(\mathbf{\theta_k})} & \ \ldots\ & \gate[6]{U(\mathbf{\theta_l})} &\\
        & & & & \ \ldots\ & &\\
        & & & & \ \ldots\ & &\\
        \myvdots \\
        & & & & \ \ldots\ & &\\
        & & & & \ \ldots\ & &
    \end{quantikz}
\end{figure}
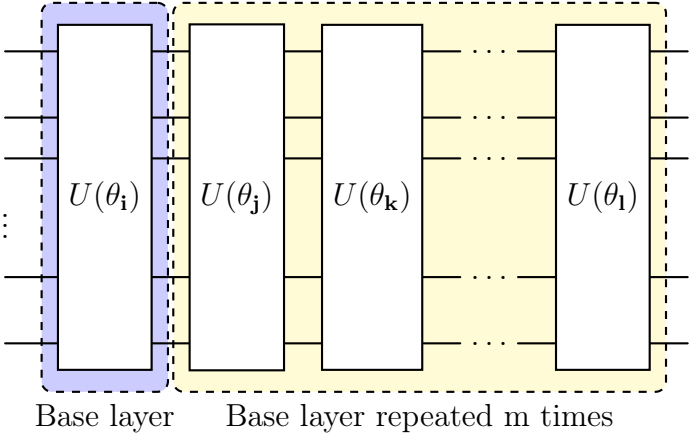

%% file: circuits/euler_parametrization.tex
\begin{figure}
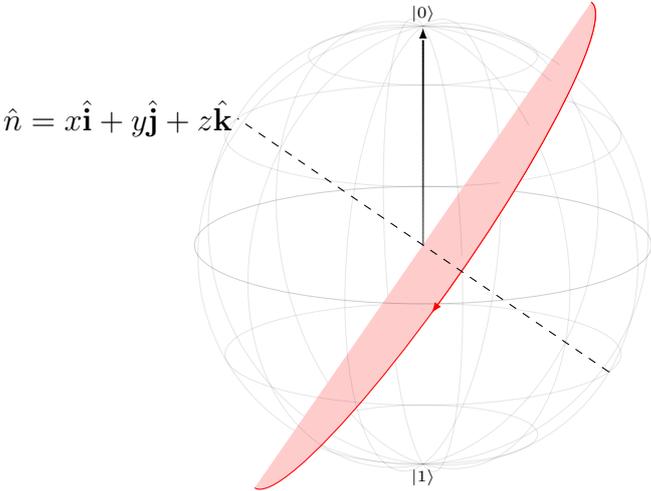

    \centering
    \caption{Euler Parametrisation in a Bloch Sphere}
    
    \begin{blochsphere}[radius=3 cm,tilt=15,rotation=-20,opacity=0]
        \drawBallGrid[style={opacity=0.1}]{30}{30}
        \labelLatLon{up}{90}{0};
        \labelLatLon{down}{-90}{90};
        \node[above] at (up) {{\tiny $|0\rangle$ }};
        \node[below] at (down) {{\tiny $|1\rangle$}};
        \drawStatePolar[]{Initial State}{0}{0}
        \drawRotationLeft[scale=1.3,style={red,fill=red!20}]{-60}{0}{0}{280}
        \drawAxis[style={dashed}]{-60}{0}
        \labelPolar[]{hat}{-60}{0}
        \node[left] at (hat){$\hat{n}=x \hat{\textbf{i}}+y\hat{ \textbf{j}}+z\hat{ \textbf{k}}$};
    \end{blochsphere}
    \label{fig:eulerparam}
\end{figure}

%% file: circuits/ep.tex
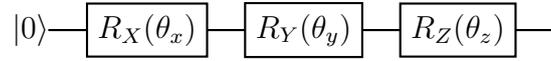
\begin{figure}[H]
    \centering
    \caption{Euler Parametrisation}
    \begin{quantikz}
        \ket{0} & \gate{R_X(\theta_x)}&\gate{R_Y(\theta_y)}&\gate{R_Z(\theta_z)}&
    \end{quantikz}
    \label{circ:ep}
\end{figure}

%% file: circuits/iqp.tex
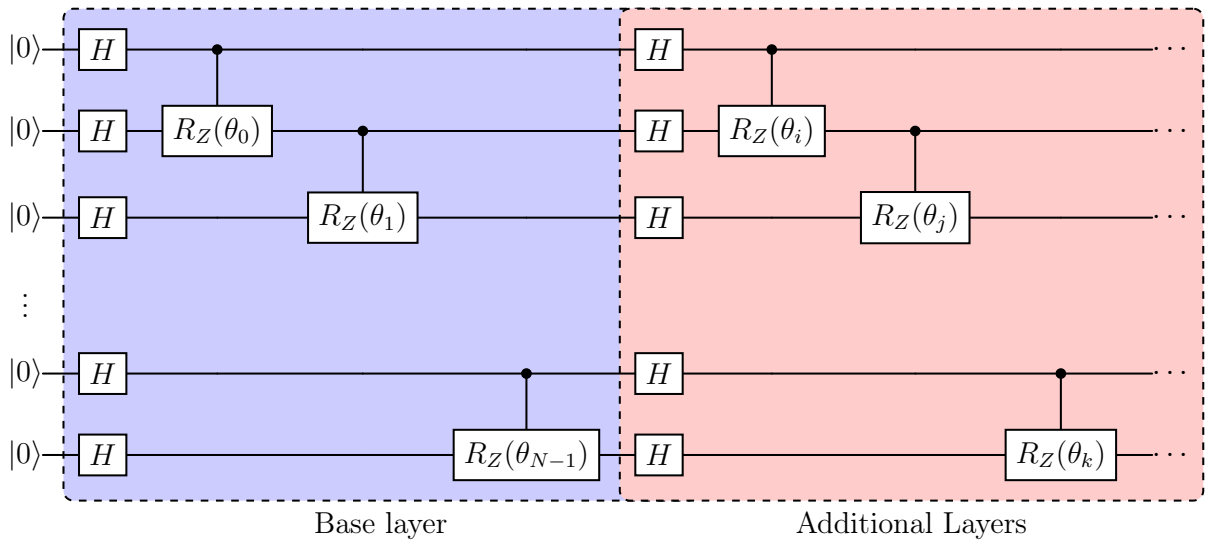
\begin{figure}
    \centering
    \caption{IQP Ansatz}
    \label{circ:IQP}
    \begin{adjustbox}{width=\textwidth}
    \begin{quantikz}
    \ket{0} & \gate{H}\gategroup[6,steps=5,style={dashed,rounded
        corners,fill=blue!20, inner
        xsep=2pt},background,label style={label
        position=below,anchor=north,yshift=-0.2cm}]{{Base layer}}&\ctrl{1}&&&\gate{H}\gategroup[6,steps=5,style={dashed,rounded
        corners,fill=red!20, inner
        xsep=2pt},background,label style={label
        position=below,anchor=north,yshift=-0.2cm}]{{Additional Layers}}&\ctrl{1}&&&\cdots\\
    \ket{0} & \gate{H}&\gate{R_Z(\theta_0)} &                               \ctrl{1} & & \gate{H} & \gate{R_Z(\theta_i)} & \ctrl{1} &&          \cdots\\
    \ket{0} & \gate{H}&&\gate{R_Z(\theta_1)} & & \gate{H} & &                \gate{R_Z(\theta_j)}& &\cdots\\
    \myvdots  \\
    \ket{0} &\gate{H}&&&\ctrl{1}&\gate{H}&&&\ctrl{1}&\cdots\\
    \ket{0} & \gate{H}&&&\gate{R_Z(\theta_{N-1})} &\gate{H}&&&\gate{R_Z(\theta_k)}&\cdots
    \end{quantikz}
    \end{adjustbox}
\end{figure}

%% file: circuits/efficient.tex
\begin{figure}[H]
    \centering
    \caption{General Hardware Efficient Ansatz}
    \begin{quantikz}
        \ket{0} & \gate{U_1}&\gate[2]{U_{\theta_1}}&&&&\gate[6,style={fill opacity=0}]{U_{\theta_N}}&\\
        \ket{0} & \gate{U_2} &&\gate[2]{U_{\theta_2}}&&&\linethrough&\\
        \ket{0} & \gate{U_3}&&&\gate[3]{U_{\theta_3}}&&\linethrough&\\
        \myvdots \\
        \ket{0} & \gate{U_{N-1}}&&&&\gate[2]{U_{\theta_{N-1}}}&\linethrough&\\
        \ket{0} & \gate{U_{N}}&&&&&&
    \end{quantikz}
    \label{circ:hea}
\end{figure}
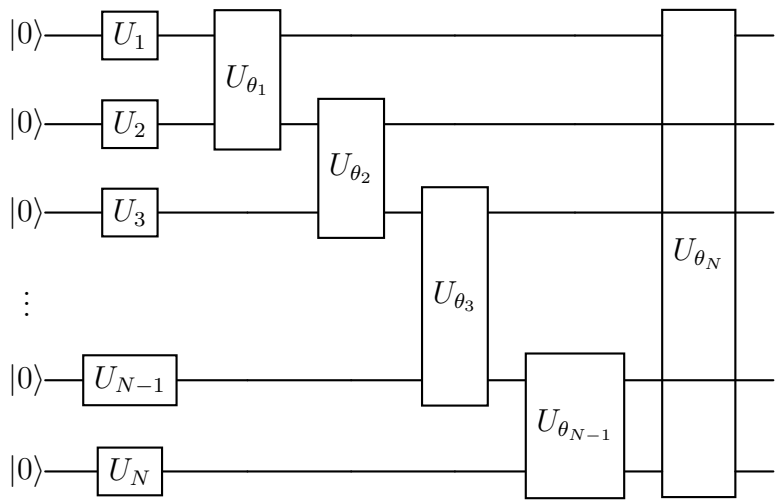

%% file: circuits/pqe.tex
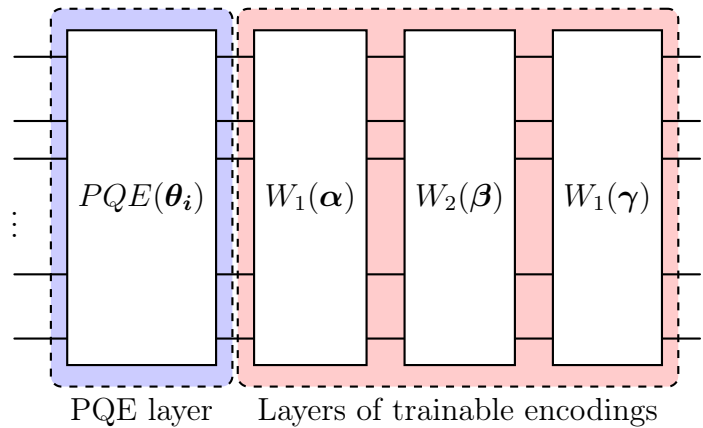
\begin{figure}[H]
    \centering
    \caption{FSL encoding scheme}
    \begin{quantikz}
        & \gate[6]{PQE(\boldsymbol{{\theta_i}})}\gategroup[6,steps=1,style={dashed,rounded corners,fill=blue!20, inner xsep=2pt},background,label style={label position=below,anchor=north,yshift=-0.2cm}]{PQE layer}&\gate[6]{W_1(\boldsymbol{\alpha})}\gategroup[6,steps=3,style={dashed,rounded corners,fill=red!20, inner xsep=2pt},background,label style={label position=below,anchor=north,yshift=-0.2cm}]{Layers of trainable encodings}&\gate[6]{W_2(\boldsymbol{\beta})}&\gate[6]{W_1(\boldsymbol{\gamma})}&\\
        & & & & &\\
        & & & & & \\
        \myvdots \\
        & & & & & \\
        & & & & &  
    \end{quantikz}
    \label{circ:pqe}
\end{figure}

%% file: 4_Problem.tex
\chapter{The Project} \label{chap:Problem}

The story so far tells us of a situation in which a method to perform useful NLP tasks on quantum computers exists, and if the NISQ era were far past it would be one that is as native to quantum systems as the simulations of quantum systems are. So DisCoCat on quantum systems is reduced to the world of variational algorithms that are costly to run and, as Shor's factorization, not useful until we can make use of bigger systems that can process vast amounts of data.

The first thought of many would be something along the lines of: "Hey, there is a direct translation between word meaning states and quantum states so why don't we train everything classically and just get the quantum state to be that of the word embedding". Indeed, in a perfect world that would be the answer, but NISQ systems strike again and for a unitary on $N$ qubits we would need around $4^N$ gates, so you say maybe we can use fewer gates and train the parameters, and then we have come full circle to the state of the art with HQCs and variational algorithms.

This reasoning is not necessarily all for nothing. During the review of the literature, we came across two interesting papers which showed some interesting possibilities \cite{nakaji_approximate_2022,mitsuda_approximate_2023}. Approximate Amplitude Encoding (AAE) as a training scheme is discussed in more depth in appendix \ref{app:aae}. It shows a method to variationally train circuits to encode in the amplitude of a quantum system some data. This idea suffers the same problems we will try to solve (which will be further explained) in this work but is an interesting path for further research.

Currently, methods for DisCoCat involve getting all of the sentences with all of the words that are going to be seen and training them on some task. For any given circuit in the testing dataset, the words it encode will already have been seen during training. OOV words cannot be inputted into the system because it will output incorrect results. This presents some obvious problems. We would have to know beforehand all of the words we want the system to learn. This is unfeasible because we cannot have perfect knowledge of the needs of the future. This will also mean training on a bigger data set, so more computation time, more resources, and more costs. And if we had a system that we would want to upgrade and add more vocabulary to, it would need to be retrained. This same problem plagues the implementation of AAE.

But what if we not only could train a circuit knowing it would work with words it hasn't seen yet, but also we could train it with a subset of the words we are aware it needs to know? This would mean that we would have found some way in which to lessen the system requirements for parameter learning and provide more flexibility to develop more complex algorithms for QNLP tasks

Given this goal and the restrictions imposed by the NISQ era, this work aims to develop an FSL framework for QNLP that will reduce the training toll required by the the NISQ era, aiming for both fast convergence to need fewer training cycles as traditional methods, as well as for reducing the impact OOV words have on a given model, thus lessening the training load on quantum systems by offloading it to classical machines.

%% file: 5_Framework.tex
\chapter{The Framework} \label{chap:framework}
What follows is our proposal for how an FSL framework for QNLP would work.

The motivation behind this framework is the word embeddings themselves. As mentioned in \ref{sec:embeddings}, the methods to obtain word embeddings usually exploit the relationship words have with each other. This structure is apparent when computing basic measures with them. Indeed, the structure that predominates is their meaning, that is what informs how the vectors are oriented in their vector space, how far apart words are, how linearly combining them can produce other words, and how their linear combination gives us a clue about the meaning decomposition. This structure is not lost in DisCoCat. The tensor product of the word strings and the linear mapping both conserve but transform it and this structure is responsible for the final form the meaning vector of a sentence takes. An FSL approach based on transfer learning that wants to make use of the abundant classical embeddings should make use of this structure since, regardless of the system type it runs on, the definition of the sentence meaning space is the same, and we could change the word space to be of type \textbf{fHilb} instead of \textbf{FVec} with little under-the-hood tinkering required. This would make the quantum state and the meaning vectors live in the same linear structure. So first, we propose that an FSL framework for QNLP should find a way to use classical embeddings to inform a variational circuit about its parameters.

So the FSL circuit should first involve an encoding which takes into account the classical embeddings. We are still left with learning the sentence space specific to the task. So the FSL circuit should then involve a variational layer that is responsible for learning the task and learning how the sentence meaning space uses the word structure.

It is apparent that by studying the necessities of our specific case, we have arrived at a variational structure similar to that described in Liu et al's paper in section \ref{sec:who}.

The general framework for encoding a word is then the following. For every word, first, a variational layer whose parameters are informed by a classical embedding of the same word is applied. This is the PQE layer and is specific for each word. That, is, the initial quantum state of a word with embedding vector $\overrightarrow{word1}$ and a word with embedding vector $\overrightarrow{word2}$ are:
\begin{align}\label{eq:PQE}
    |\psi_1\rangle&=U(\{ \boldsymbol{\theta}_i \})|0\rangle \\
    |\psi_2\rangle&=U(\{ \boldsymbol{\varphi}_j \})|0\rangle,
\end{align}
respectively. Where $|\psi_1\rangle=|\psi_2\rangle \iff \overrightarrow{word1}=\overrightarrow{word2}$. This ensures that the mapping is unique. The mapping should also be deterministic. There can be a variety of ways to choose a mapping $A:\overrightarrow{word}\rightarrow \{\boldsymbol{\theta}_i\}$ and it does not necessarily have to be linear. This mapping should be informed mostly by the structure from the embedding set that is wished to be preserved (e.g. the inner product among word embeddings). This layer is fixed and does not change through the training epochs. The mapping $A$ should be calculated in a classical system.

The next step is to construct a variational layer after the PQE that will be in charge of learning the task. The key step for this layer is that for words of the same pregroup type, the unitary applied to the quantum state should be the same. That is for two words $\overrightarrow{word1}$ and $\overrightarrow{word2}$, their complete encoding should be:
\begin{align*}
    |\Psi_1\rangle&=W(\{ \boldsymbol{\alpha_i} \})|\psi_1\rangle=W(\{ \boldsymbol{\alpha_i} \})U(\{ \boldsymbol{\theta}_i \})|0\rangle \\
    |\Psi_2\rangle&=W(\{ \boldsymbol{\beta_j} \})|\psi_2\rangle=W(\{ \boldsymbol{\beta_j} \})U(\{ \boldsymbol{\varphi}_j\})|0\rangle,
\end{align*}
where $W(\{\boldsymbol{\alpha}_i\})=W(\{ \boldsymbol{\beta}_i \}) \iff P.type(\overrightarrow{word1})=P.type(\overrightarrow{word2})$. That the same $W$ is used for all words of the same pregroup type means that the model can properly learn how the structure of the words affects the model output, instead of the PQE being just a costly state initialisation state.

A few considerations exist for choosing $W$. It does not need to be that the variational circuit for $W$ should be the same for all pregroup types. One can choose one ansatz for Nouns and another for Sentences, but it is easier to allow for the same base ansatz in all cases (although further research could be done to assert whether this comes with a trade-off in performance). The expressibility should also be considered. This was not mentioned for the PQE layer because the structure worth preserving usually leaves in a lower dimensional space than the $2^N$ dimension Hilbert space for $N$ qubits. The inner product (or cosine similarity) is just a single number for two embeddings regardless of the number of qubits or vector dimensions. Even if the PQE were severely limited in expressibility, the higher-dimensional manifold could admit a lower-dimensional projection of the structure-preserving solution. Furthermore, the output of the PQE is just another quantum state. Since it is fixed for all epochs, it does not add nor subtract from the expressibility of $W$. The $W$ layer should be responsible for finding a solution and so is more susceptible to the trappings of expressibility. The PQE would affect the solution by affecting the subset of the search space available to $W$, but this would be mitigated by choosing a relatively highly expressible ansatz. In $W$ is also where practical considerations like qubit architecture should be considered.

Figure \ref{circ:fslex} shows what a circuit implementing this framework would look like. This circuit is for a transitive sentence, where we remember that if a noun has type $N$, we map it to a two-qubit state, and the transitive verb has type $n(n^rsn^l)n$ and it is mapped to a five-qubit state. It is readily apparent how both words of type $N$ share a common variational layer $W$, different than that for the transitive verb. It is also apparent how even though two words of the same type $N$ appear in the sentence, both have a different PQE parametrization. The quantum state which encodes the meaning of the sentences is the middle wire, and we only consider only those output states where the measured qubits gave a measurement of 0. 

\input{circuits/fslqex}

The process to train a circuit using this framework would be the following:
\begin{enumerate}
    \item Define both a classical embedding set and a structure which will be preserved in the quantum circuit.
    \item Define the ansatz scheme that will be used for both the PQE and the variational layers, and for each pregroup type.
    \item Construct a deterministic mapping that will take as input the classical embedding of a word and will injectively output a set of parameters for the PQE layer.
    \item Run the training cycle as normal, updating the shared parameters in $W$ according to a cost function, evaluating the circuits accordingly, and sharing the processing among quantum and classical systems.
\end{enumerate}
After training a circuit, the deterministic mapping should take in a word that the model has not seen before and output a parametrization that, when run through the model, gives an output that is coherent with the rest of the words seen during training.

%% file: circuits/fslqex.tex
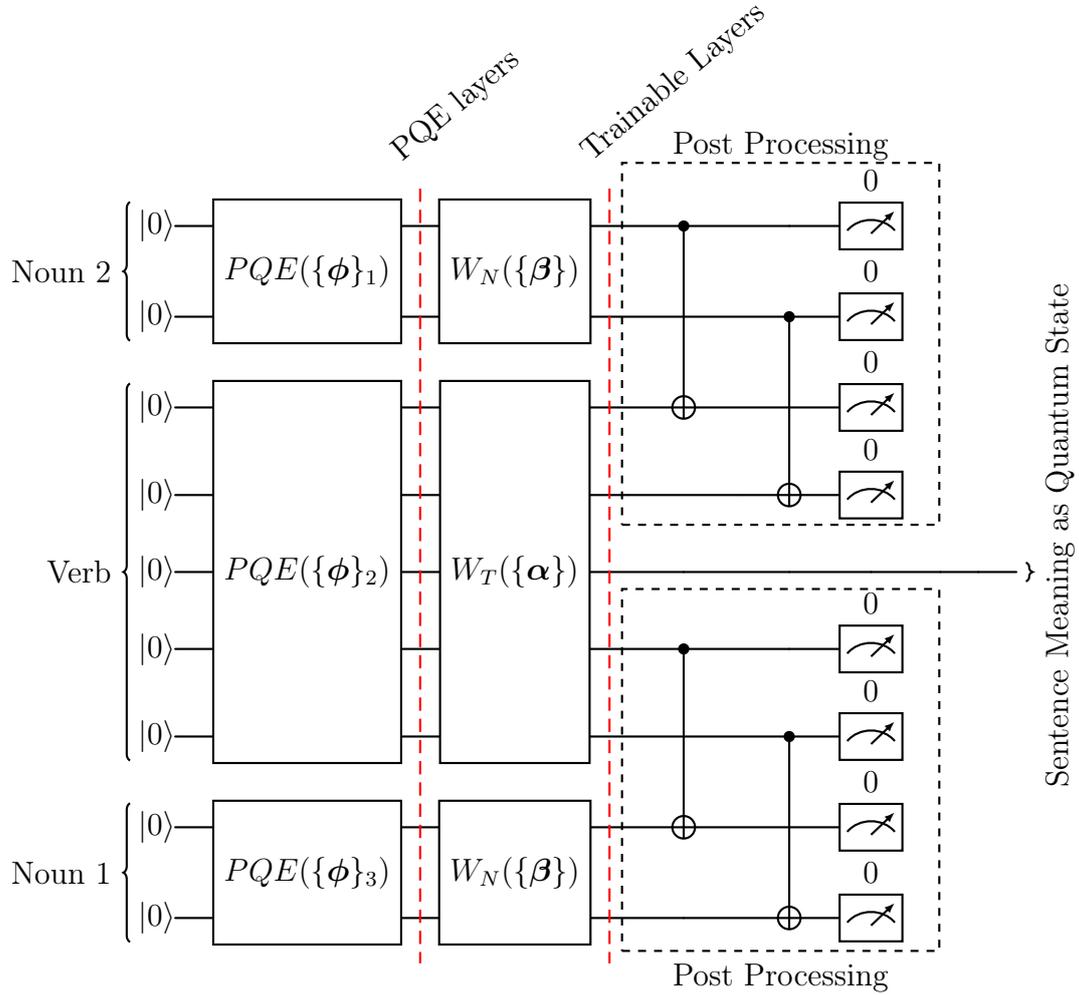
\begin{figure}[H]
    \centering
    \caption{FSL scheme in a circuit}
    \begin{quantikz}
        \lstick[2]{Noun 2}\ket{0}&\gate[2]{PQE(\{\boldsymbol{\phi}\}_1)}\slice[label style={inner sep=1pt,anchor=south west,rotate=40,yshift=0.3cm}]{PQE layers} &\gate[2]{W_N(\{\boldsymbol{\beta}\})} \slice[label style={inner sep=1pt,anchor=south west,rotate=40,yshift=0.3cm}]{Trainable Layers}& \ctrl{2} \gategroup[4,steps=3,style={dashed,xshift=0.2cm,yshift=0.2cm}]{Post Processing} & & \meter{0}\\
        \ket{0}& & & &\ctrl{2} & \meter{0}\\
        \lstick[5]{Verb}\ket{0}&\gate[5]{PQE(\{\boldsymbol{\phi}\}_2)}&\gate[5]{W_T(\{\boldsymbol{\alpha}\})} & \targ{} & & \meter{0}\\
        \ket{0}& & &  \linethrough &\targ{}&\meter{0}\\
        \ket{0}& & & & & & & &\rstick[2,label style={rotate=90,yshift=-0.4cm,xshift=-3cm}]{Sentence Meaning as Quantum State}\\
        \ket{0}& & & \ctrl{2}\gategroup[4,steps=3,style={dashed,xshift=0.2cm,yshift=0.2cm},label style={label
position=below,yshift=-0.55cm}]{Post Processing}& & \meter{0}\\
       \ket{0}& & & & \ctrl{2}& \meter{0}\\
         \lstick[2]{Noun 1}\ket{0}& \gate[2]{PQE(\{\boldsymbol{\phi}\}_3)}&\gate[2]{W_N(\{\boldsymbol{\beta}\})} & \targ{}& & \meter{0}\\
        \ket{0}& & & & \targ{} & \meter{0} 
    \end{quantikz}
    \label{circ:fslex}
\end{figure}

%% file: 6_Experiment.tex
\chapter{Framework Testing}\label{chap:experiment}
\section{The Task}
To test this framework, a series of different circuits will be tested in a meaning classification task (MC).

In this task, we present the circuit with a sentence and ask it to tell us if the sentence talks about food or technology. Example sentences from the training set can be seen in table \ref{tab:examples}.

\begin{table}[H]
\centering
\caption{}
\begin{tabular}{|l|l|c|ll}
\cline{1-3}
\multicolumn{1}{|c|}{\textbf{Sentence}} & \multicolumn{1}{c|}{\textbf{Theme}} & \textbf{Label} &  &  \\ \cline{1-3}
Woman cooks flavorful soup.             & Food                                & 1              &  &  \\ \cline{1-3}
Man makes useful application.           & Technology                          & 0              &  &  \\ \cline{1-3}
Man makes supper.                       & Food                                & 1              &  &  \\ \cline{1-3}
Guy debugs helpful application.         & Technology                          & 0              &  &  \\ \cline{1-3}
\end{tabular}
\label{tab:examples}
\end{table}

Within this task, we define the sentence space to be a 2 dimensional Hilbert space whose canonical basis $\hat{e}_1=[1,0]$, $\hat{e}_1=[0,1]$ is mapped by construction to the sentence labels. So our meaning space is a 2 dimensional Hilbert space spanned by the concepts of technology and food.

\section{The \mya}
\subsection{Traditional Benchmark}
As a control and to get a baseline for how the traditional \mya would perform, two were chosen: IQP ansatz and Sim15Ansatz (figure \ref{circ:Sim15}). The IQP ansatz was previously discussed in section \ref{subsec:ansatz}, and was chosen for both its simplicity, in terms of trainable parameters and cost to simulate. Sim15Ansatz was chosen because it has a higher expressibility than IQP, while still being efficient to simulate.

Nothing was changed with how the ansatz would normally be declared, implemented, and run.

\input{circuits/sim15}

\subsection{FSL \mya}
The definition of the \mya changes when implementing the FSL scheme. The \mya consists of two parts, the PQE, and the variational $W$ layer.
\subsubsection{Pre Quantum Embeddings}
To test this framework we decided on two ways to deterministically calculate the PQE based on some set of classical embeddings. 

The classical embeddings chosen for this experiment were Stanford's GloVe embeddings, specifically the 300-dimensional Common Crawl version. This was chosen because it is a co-occurrence unsupervised algorithm which necessitates less computational power than other models such as BERT. It has enough dimensions to provide a qualitatively complex enough representation of the data without being too costly to manage.

The first of the two methods is called the \textbf{naive approach}, further referred to as \textbf{FSL Base}. This approach comes out of the need to preserve some structure of the data while also being conscious that anything other than a 1 qubit state is costly to perfectly create. So if we picture the embedding as a vector in N-dimensional space, to map it into a single qubit state we could reduce the dimensionality to 3, and thus obtain the rotation angles needed for the Euler parametrisation. This would conserve the structure of the reduced-dimension vector space.

We call this the naive method because of two things. First, it is impossible to preserve all of the inherent structure when reducing the dimensions, so if a "structure-preserving map" is sought, this might be lacking. This method also requires the reduced dimensionality to be 3, because we run with the same $4^N$ gates cost when trying to make an N qubit state map to a $2^N$ vector space. What happens, then, when we need a Hilbert corresponding to more than one qubit? Each qubit is initialised with the same Euler parametrisation (figure \ref{circ:naive}).

\input{circuits/naivepqe}

It is reasonable to ask precisely what structure is preserved. If we assume the relationship between words is preserved in the inner product of the dimension-reduced vectors, then we can draw the following conclusion. 

By how we construct the quantum states, this holds: $\langle \psi_{word_1}|\psi_{word_2}\rangle = \overrightarrow{word1} \cdot \overrightarrow{word2}$=C, and for a multi-qubit state:
\begin{align*}
    |\psi_{word}\rangle &= U(word)^{\bigotimes N}|0\rangle^{\bigotimes N},\\
    |\psi_{word}\rangle &= | word \rangle ^{\otimes N},
\end{align*}
where we extended equation \ref{eq:PQE} to $N$ separable states and $ | word \rangle ^{\otimes N}=U(word)^{\bigotimes N}|0\rangle^{\bigotimes N}$. It is then straightforward to arrive at the following conclusion,
\begin{align}\label{eq:struc}
    \langle \psi_{word_1}|\psi_{word_2} \rangle &=\otimes_{i=0}^N \langle word_1|word_2 \rangle_i, \\
    \langle \psi_{word_1}|\psi_{word_2} \rangle &= C^N.
\end{align}
What equation \ref{eq:struc} tells us is that the inner product could be considered somewhat preserved. This naive PQE amplifies the inner product of words closer together while decreasing that of words farther away.

For our experiment, we choose to use the t-SNE dimensionality reduction method included in the Scikit package. This method was chosen as it includes a deterministic version of the algorithm. This is important since we want the PQE for each word to not change on different calculations of the parameters.

The second method we call \textbf{FSL NN}, as it is based on a neural network. Following the idea of preserving the inner product of the word embeddings, the main idea of this method is to train a neural network that takes as input the embedding of a word and outputs the rotation angles for a given ansatz. 

This neural network was constructed and trained completely classically, using PyTorch to create and train it. The cost function was defined to be the MSE between the inner product squared of the two embeddings and the fidelity of the quantum states whose parameters correspond to the output of the neural network.

A PQE model for \textbf{NN} method would consist of the PyTorch NN model. To get the PQE parametrization, the word vector is fed to the NN, then the output parameters are used in the PQC. To train the NN the training samples consisted of all the words, seen and unseen for quantum training. The labels are just the inner product squared of the vector embeddings.

The PQC chosen is circuit 4 (figure \ref{circ:NN}) in \cite{sim_expressibility_2019} (with a Ry gate instead of the Rz gate). This circuit was chosen because it has a relatively high expressibility for a low number of gates. We want the NN to find the optimal quantum state that has the same inner product structure as the vector embeddings, so letting it explore a higher portion of the Hilbert Space is of benefit.

\input{circuits/NN}

\subsubsection{Variational Layer}

The variational layer is the circuit 4 from \cite{sim_expressibility_2019} without any changes. Apart from having a high expressibility, its expressibility also increases substantially with additional layers for the first two, so we can test if subsequent layers help with finding an optimal solution. Circuit 4, can also be seen to be a specific implementation of a HEA. Thus adding a benefit for its implementation on real devices.

The variational layer has the same structure for all pregroup types, but a useful reminder is that even though the building structure is the same, different pregroup types have to have different parameters.

\section{Training tools and training flow}

To train and test the model, the Lambeq Python package was used. The training flow consisted of first reading all the sentences from the text files and converting them to a data-label pair. The sentences were then tokenised using the included tokeniser module (i.e. split into their constituent words). These tokens were parsed using the Bobcat parser and turned into their corresponding DisCoCat diagrams. The circuit ansatz is then chosen and each diagram is converted to its corresponding quantum circuit.

After all the sentences were turned into a circuit, training begins. Lambeq takes care of evaluating each circuit and post-processing. Since the sentence space is two-dimensional, then the output of the circuit is a single qubit, and evaluation gives as a result a two-dimensional normalised vector, which is then rounded by convention to its nearest basis state (e.g. if the output were $[\sqrt{0.75},\sqrt{0.25}]$ then the model final prediction would be $[1,0]$). We have defined the prediction to be 0 if the output is $[1,0]$ and 1 if it is $[0,1]$. 

We used the included Binary Cross Entropy as a loss function, and accuracy was calculated to be the percentage of successful classifications from the total. We also used the included SPSA optimiser. We used a Quantum Trainer (as it is the trainer that works for quantum circuits) and the model is the NumpyModel. We chose this model because we are simulating the quantum circuits, so this allows the simulation to use fewer computational resources. 

The following are the training parameters used:
\begin{itemize}
    \item \textbf{a} : 0.05
    \item \textbf{c} : 0.06
    \item \textbf{A} : 0.01*Epochs
    \item \textbf{Epochs Behavioural Testing} : 2000
    \item \textbf{Epochs OOV Testing} : 1500
    \item \textbf{Batch size} : 700
\end{itemize}
To account for randomness, the training cycle was repeated with each of the following initialisation seeds: $[0, 10, 50, 77, 100, 111, 150, 169, 200, 234, 250, 300, 350, 400, 450]$. 

Due to the particular details of Lambeq, the models had to be initialised before training with all of the circuits that would be evaluated at some point or another. So the original initialisation was done with the OOV sentences as well. Nevertheless, they were not shown during training and training does not impact the OOV parameters. This amounts to essentially randomising the parameters.


\section{Behavioural Testing}

To first obtain some data on the training behaviour of this framework, the MC task was implemented using the FSL Base PQE, and compared with the Sim15 ansatz. These were chosen because they are a reliable representation of their classes (e.g. the traditional and the new FSL framework) while also being computationally inexpensive to calculate. 

This testing aims to see if training would be effective with fewer parameters and a shared structure and if the training curve could be different. Links to all code can be found in appendix \ref{System Requirements}.

Testing was done on circuit sizes [2,3,4,5], with a second layer also tested for circuit sizes [2,3,4].

\subsection{Datasets}
The dataset for this section is obtained from the Lambeq documentation for MC \cite{kartsaklis_lambeq_2021}. It consists of:
\begin{itemize}
    \item \textbf{Training Set:} 67 sentences
    \item  \textbf{Development Set:} 28 sentences
    \item \textbf{Test Set:} 29 sentences
\end{itemize}

\subsection{Results}
\subsubsection{Convergence and Size Dependence}
Training behaviour can be seen graphically in figure \ref{fig:test1}, and these give light on how the FSL framework compares to traditional \mya.

The circuit width changes the behaviour of \mya when compared to each other. At low-width circuits, the Sim15 ansatz outperforms the FSL Base both in convergence and accuracy, but when the circuit gets wider, the FSL Base outperforms its traditional counterpart. At the width of 2 qubits, FSL Base converges in approximately half the epochs of Sim15. By widths 3 and 4, the accuracies start to be similar, and at width 5 the accuracy of the FSL Base exceeds that of the Sim15.

When looking at the circuit depth, a similar behaviour occurs. Since we want to consider only shallow circuits, the depth should be, in general, considerably smaller than the width. For depths 2 at widths 3 and 4, it is evident that Sim15 converges linearly and does not reach the same accuracy level as the FSL Base implementation.

This behaviour is attributed to the increasing number of training parameters a given ansatz has to learn during training. For a traditional ansatz, each word has its own set of parameters and these grow with the size of the circuit. For a word of width N and depth L, the number of parameters a circuit has to learn is $P_{Sim15}(N,L)=L*N(P.type)$. N has to change according to the functorial definitions used, but assuming all words are of the same width, the amount of parameters learned for M words is $P_{Sim15}(N,L,M)=M*L*N$. For \mya of the HAE type, this becomes $P_{HAE}(N,L,M)=M*L*(3N-1)$. However, when we consider an ansatz of the FSL type, we only have to train a fixed number of parameters for any set of words of the same pregroup type. So for a variational layer of the form used in this project, $P_{FSL}(N,L,W)=3N-1$ is not dependent on the number of words. Both frameworks scale linearly with circuit width and depth, but for even a small dataset, the prefactor added by having to train a unique set of parameters for each word adds strain to the training process. The formulation of FSL in this work gives an optimisation to the number of parameters the model has to learn, making it a constant regardless of the number of words in the dataset (with the caveat that this remains constant among all pregroup types). In the NISQ era where every bit of performance matters, this optimisation helps.

The parameter design also helps explain why Sim15 outperforms the FSL Base for smaller circuits. Since Sim15 learns the parameters for each single word, when it has to learn a few parameters it just learns the perfect parameterization for each word. But when circuits are bigger, it doesn't have enough time to learn all of the parameters for all of the words, so its performance falls. The opposite happens in FSL, since it has to find a transformation that satisfies the task requirements for all the words in the dataset, for a few parameters it struggles to find the optimal rotation set, but when the circuits get bigger it doesn't suffer the drawbacks suffered by traditional \mya and can leverage this higher dimensional space to find a better solution. 

\input{figures/test1}

\subsubsection{Spread}

It is also worth looking at the spread of the training behaviour for different seeds (figure \ref{fig:testspread}). The accuracy large shift from epoch to epoch in Sim15 is expected, as the number of parameters that have to change is higher than in the FSL Base implementation. It is also notable how the initial randomness does not have that much of an impact in FSL Base as in Sim15, as the variation when training wider circuits, remains the same as that in smaller circuits, a behaviour which doesn't occur in the traditional Sim15 implementation.
\input{figures/tesspread}

\section{OOV Accuracy Testing}
After validating FSL's ability to learn and optimised training behaviour, we want to test the OOV learning capacity of the framework.

To test this, we choose a prequantum embedding that maximises structure preservation, so we will test NN FSL, along the other \mya.

\subsection{Datasets}

To augment the data used in the previous section we ran the individual word strings through the GloVe embeddings dataset to obtain a list of words that are closest in meaning. From each word, we selected between two and three words. These words are not necessarily those closest in meaning to the original words, as some diversity is sought. After these new words were selected, new sentences were created from the permutations of these words and nonsensical sentences were manually cleaned. With this new set of sentences five new datasets were created:

\begin{itemize}
    \item \textbf{Training Set:} 1523 sentences.
    \item \textbf{Development Set:} 522 sentences. All words in the sentences have been seen during training.
    \item \textbf{Test Set:} 509 sentences. All words in the sentences have been seen during training.
    \item \textbf{Redundancy Test Set:} 1263 sentences. Around half of the words have been seen during training and half haven't been.
    \item \textbf{OOV Test Set:} 73 sentences. All words are unseen.
\end{itemize}

This splitting has been done to be able to test the models both traditionally, with words all seen during training, and assess their learning performance, and their OOV performance.

Given the vastly augmented datasets, and the computational time and resources needed to run the simulations, a reduced set of sizes was selected, considering only circuits of size 2, 3, and 4, of a single layer. The \mya chosen to test this is the FSL Base, FSL NN, and IQP. Due to the particular behaviour seen in previous testing of NN FSL, the epochs were reduced to 1500. The training and testing choices were made to obtain a representative example of the behaviour of FSL considering the resource constraints.

\subsection{Results}
\subsubsection{Accuracy}

The testing accuracy for the words seen during training, a mix of words seen during training and unseen words, and OOV words can be seen in table \ref{tab:accuracies}. These accuracies tell us something about the behaviour of FSL in QNLP. First of all, it is noticeable how even though for smaller training sets the higher the circuit size the better FSL performs, for datasets of sizes like this one, it performs worse on width 4 than on 3, but more will be said in the next subsection.

If we look at the columns for OOV words, the accuracies for both FSL Base and IQP remain around 50\%, which is consistent with random guessing. So the conclusion is that for circuits and datasets of this form, the FSL Base prequantum embedding is only useful for early convergence, and not for OOV words.

The same is true for the mixed seen and unseen words column. We expect the accuracy to be a combination of the OOV and seen columns, which seems to be the case, as the accuracy is, in general, halfway between both of those accuracies. Even though for FSL NN the OOV accuracy is higher, it takes a penalty for the low accuracy in the seen column and thus has much lower accuracy. As for FSL Base, FSL NN has its environment in which it should be used since it is what it is designed for. 

\input{tables/accuracies}

\subsubsection{Behaviour}
A few things become apparent by looking at the behaviour of the three models (figure \ref{fig:2spread}) on a larger dataset. FSL NN converges more quickly than IQP (and by extension traditional \mya) and FSL Base. Nevertheless, this convergence comes at the cost of accuracy. FSL NN also becomes unstable at low circuit sizes, this is attributed as in the previous section to the low count of trainable parameters and thus the high impact each of them has.

When crossing the threshold to higher circuit sizes for a dataset of this size, traditional \mya might exhibit superior accuracy when compared to FSL, in comparison with lower circuit sizes. This could be explained by the fact that FSL is geared to work in a limited-resource environment. When constructing a dataset of this size, the number of individual words is not greatly increased. In this case, it increased by a factor of around five. Since the sentences that form each training point are composed of multiple words, and different permutations of these words are also valid sentences within this training set, then many more sentences are created, which contradicts the principal tenet of FSL, low resource availability.

\input{figures/2test}

However, even if we don't see a superior accuracy for seen words, the fast convergence is seen in both versions of the FSL prequantum embedding, being more prominent in FSL NN. FSL NN outperforms FSL Base because the structure finding is tasked to the classical system in a more extensive way in the former than the latter, so more training cycles are required to reach the same level of convergence in FSL Base, even if the classical system has helped it.

%% file: circuits/sim15.tex
\begin{figure}
    \centering
    \caption{Single layer Sim15 Ansatz}
    \label{circ:Sim15}
    \begin{quantikz}
        & \gate{R_y}&\targ{}&\ctrl{1}&&&\\
        &\gate{R_y}&&\targ{}&\ctrl{1}&&\\
        &\gate{R_y}&&&\targ{}&&\\
        \myvdots \\
        &\gate{R_y}&&&&\ctrl{1}&\\
        &\gate{R_y}&\ctrl{-5}&&&\targ{}&
    \end{quantikz}
\end{figure}
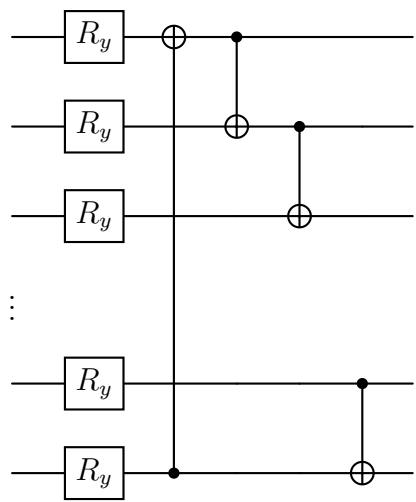

%% file: circuits/naivepqe.tex
\begin{figure}[H]
    \centering
    \caption{Base PQE}
    \begin{quantikz}
        \ket{0} & \gate{R_x(\theta_x)} &\gate{R_y(\theta_y)} &\gate{R_z(\theta_z)} &\gate[6]{W layer}&\\
        \ket{0} & \gate{R_x(\theta_x)}&\gate{R_y(\theta_y)} &\gate{R_z(\theta_z)} &&\\
        \ket{0} & \gate{R_x(\theta_x)}&\gate{R_y(\theta_y)} &\gate{R_z(\theta_z)} &&\\
        \myvdots \\
        \ket{0}&\gate{R_x(\theta_x)}&\gate{R_y(\theta_y)} &\gate{R_z(\theta_z)} &&\\
        \ket{0} &\gate{R_x(\theta_x)}& \gate{R_y(\theta_y)} & \gate{R_z(\theta_z)} & &    
    \end{quantikz}
    \label{circ:naive}
\end{figure}

%% file: circuits/NN.tex
\begin{figure}
    \centering
    \caption{PQE ansatz}
    \label{circ:NN}
    \begin{quantikz}
        \ket{0} & \gate{R_x}&\gate{R_y}& \ctrl{1}&&&\\
        \ket{0} & \gate{R_x}&\gate{R_y}&\gate{R_x}&\ctrl{1}&&\\
        \ket{0} & \gate{R_x}&\gate{R_y}&&\gate{R_x}&&\\
        \myvdots \\
        \ket{0} & \gate{R_x}&\gate{R_y}&&&\ctrl{1}&\\
        \ket{0} & \gate{R_x}&\gate{R_y}&&&\gate{R_x}&
    \end{quantikz}
\end{figure}
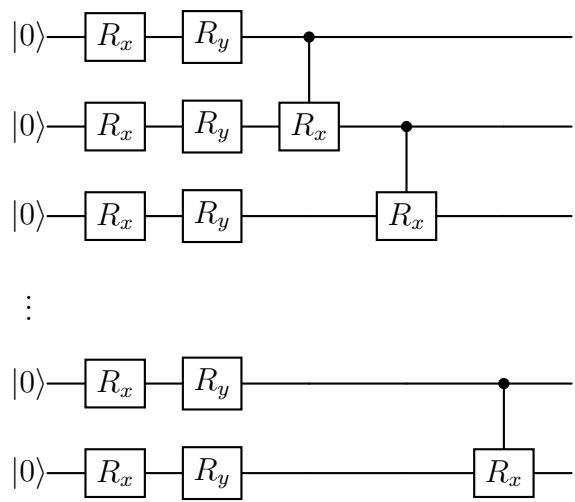

%% file: figures/test1.tex
\begin{figure}[ht]
\centering
\includegraphics[width=0.49\textwidth]{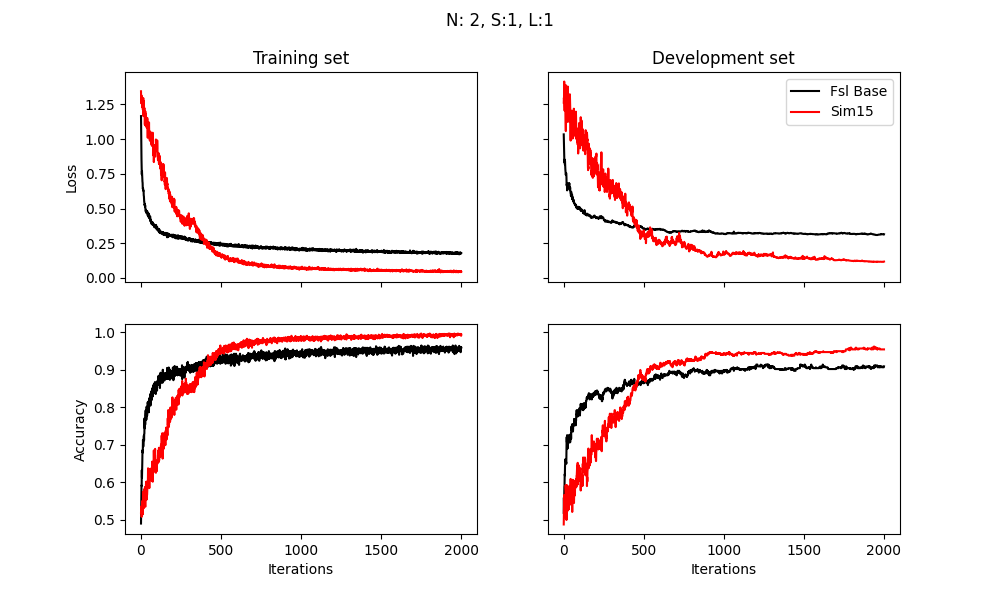}
\includegraphics[width=0.49\textwidth]{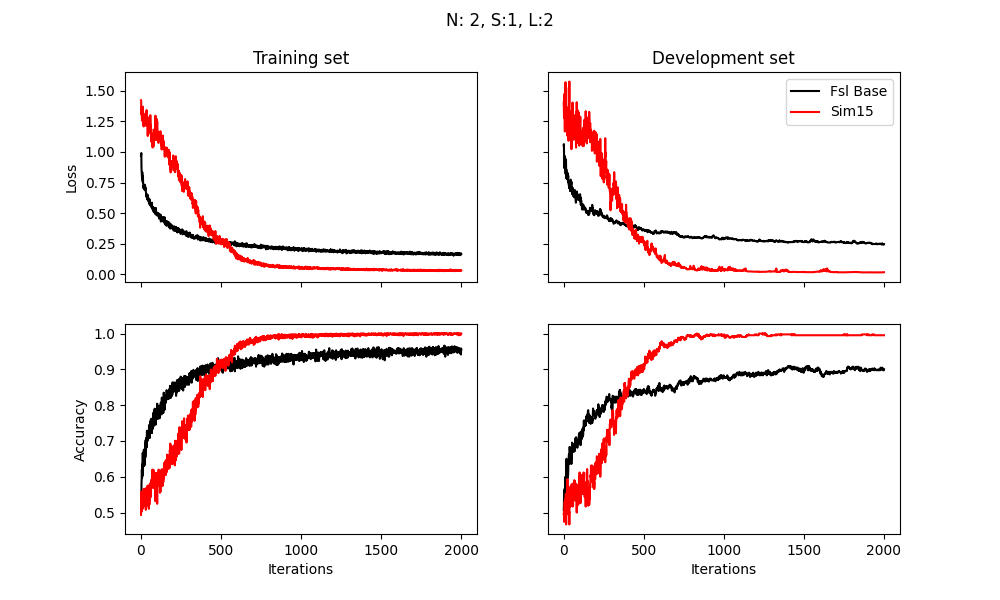}
\includegraphics[width=0.49\textwidth]{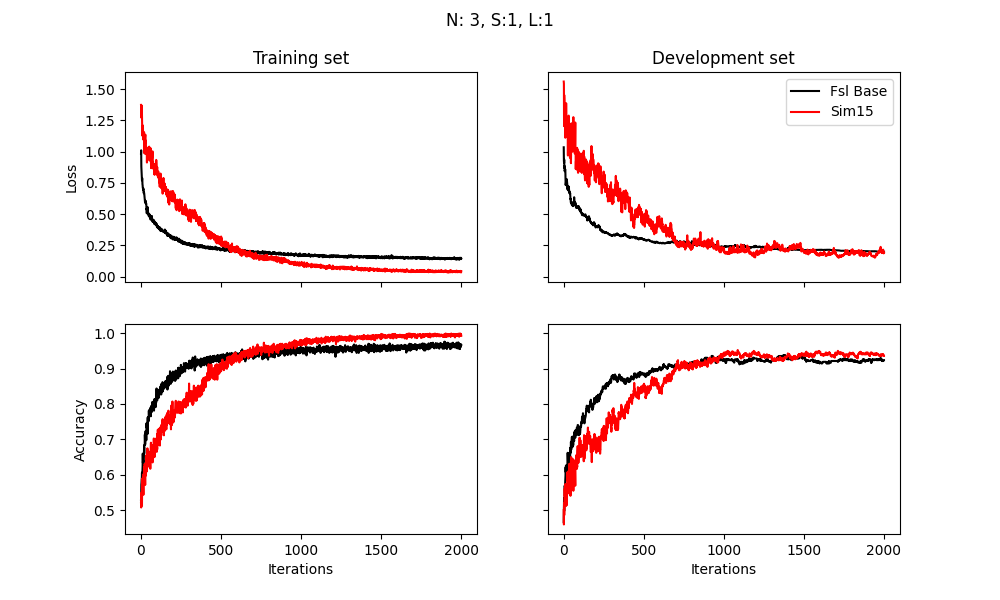}
\includegraphics[width=0.49\textwidth]{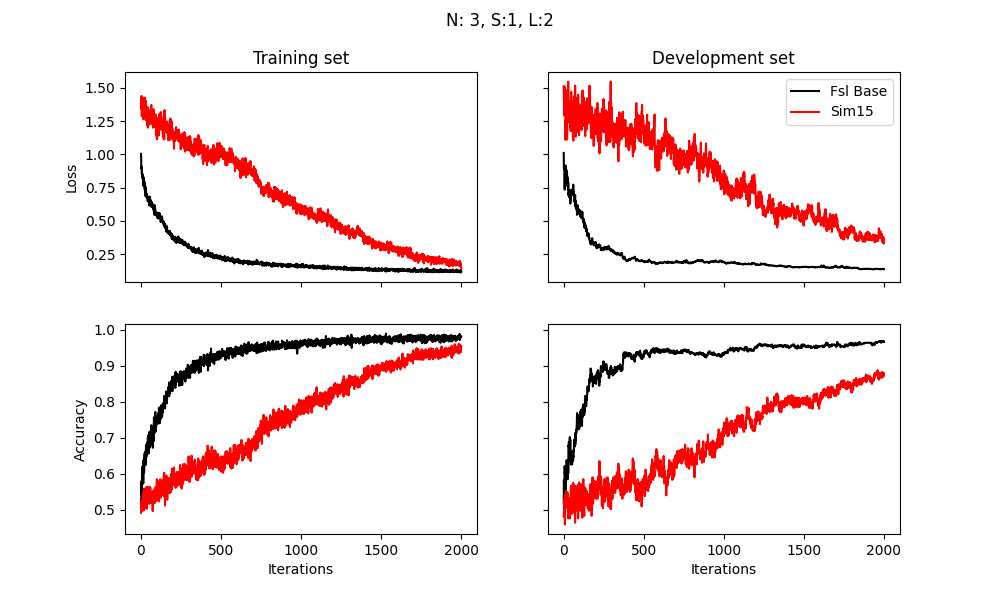}
\includegraphics[width=0.49\textwidth]{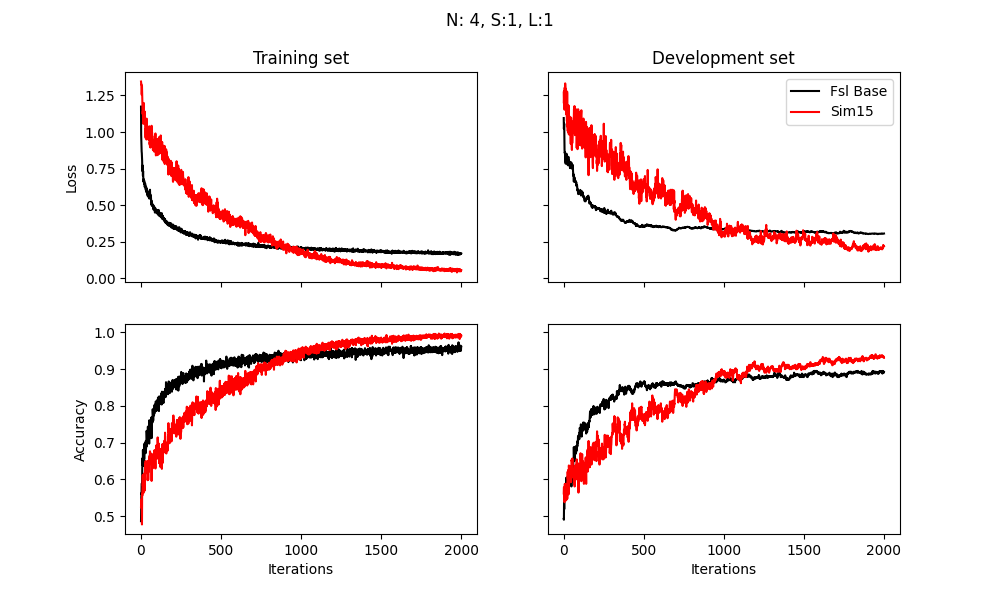}
\includegraphics[width=0.49\textwidth]{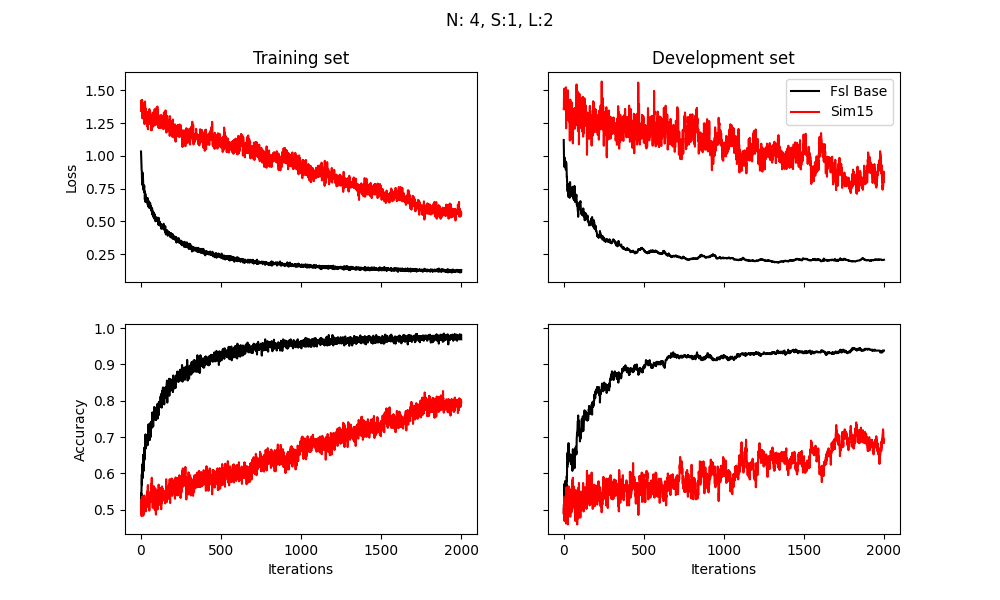}
\raggedright
\includegraphics[width=0.49\textwidth]{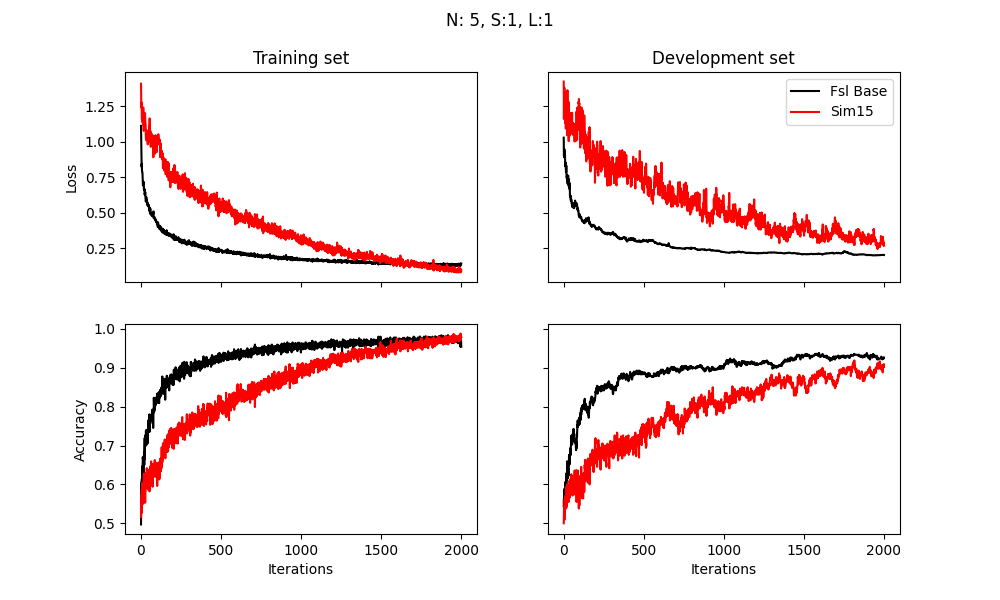}
\caption[Ans{\"a}tze behaviour comparison for small datasets]{Training behaviour for two types of \mya on the MC task after 2000 training Epochs.}
\label{fig:test1}
\end{figure}

%% file: figures/tesspread.tex
\begin{figure}[h]%
\centering
\includegraphics[width=0.49\textwidth]{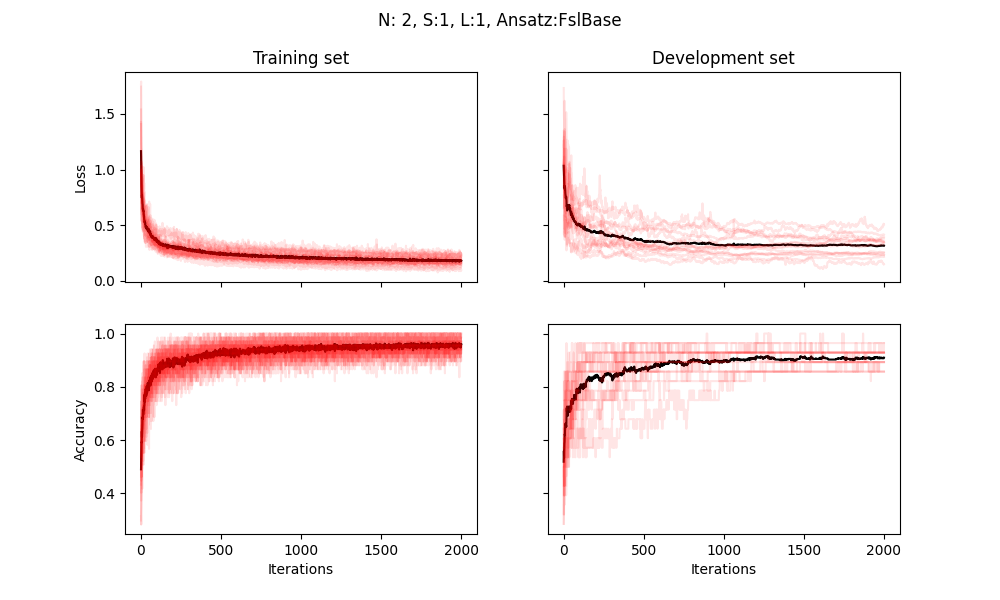}
\includegraphics[width=0.49\textwidth]{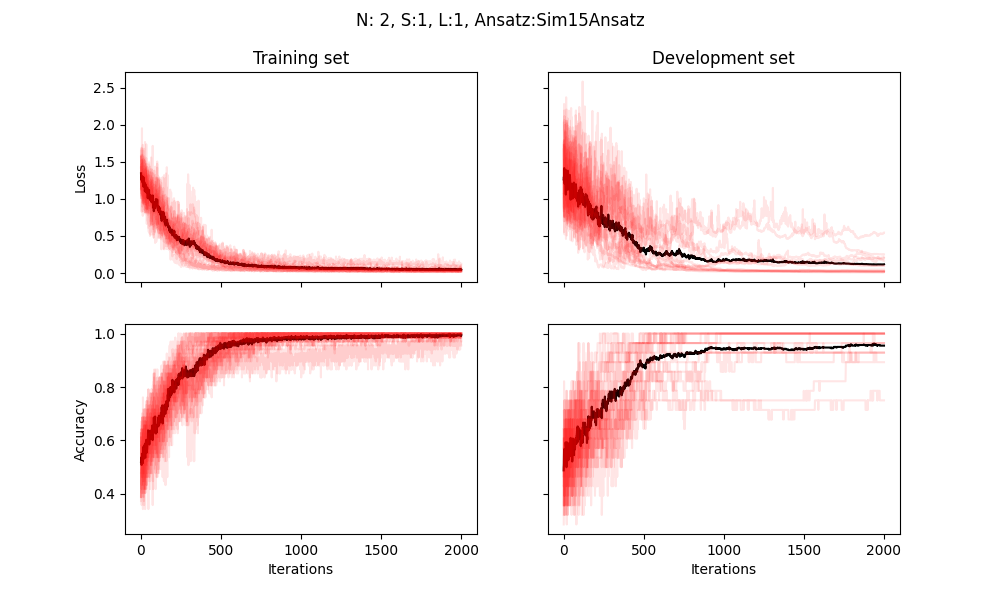}
\includegraphics[width=0.49\textwidth]{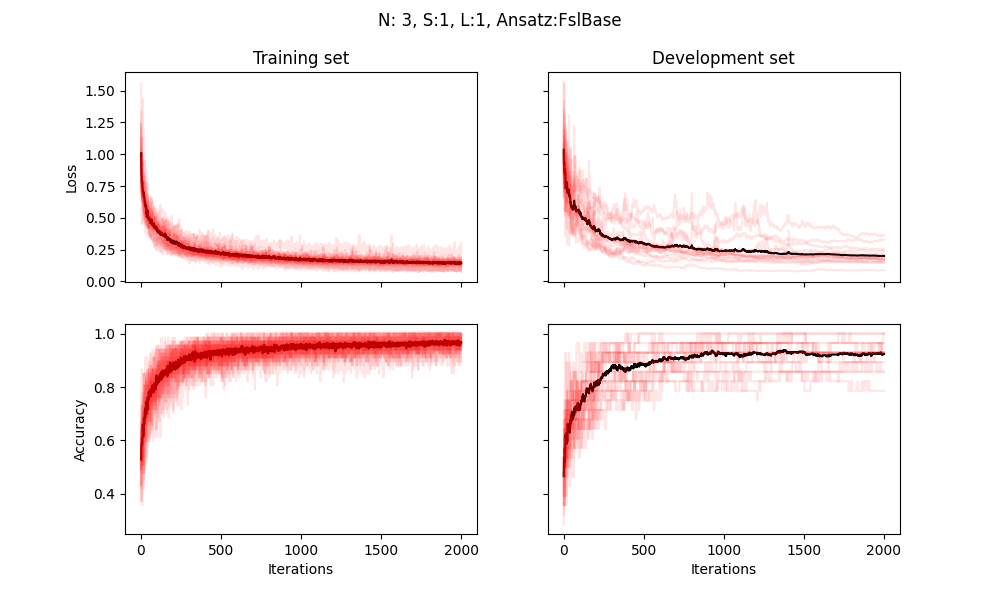}
\includegraphics[width=0.49\textwidth]{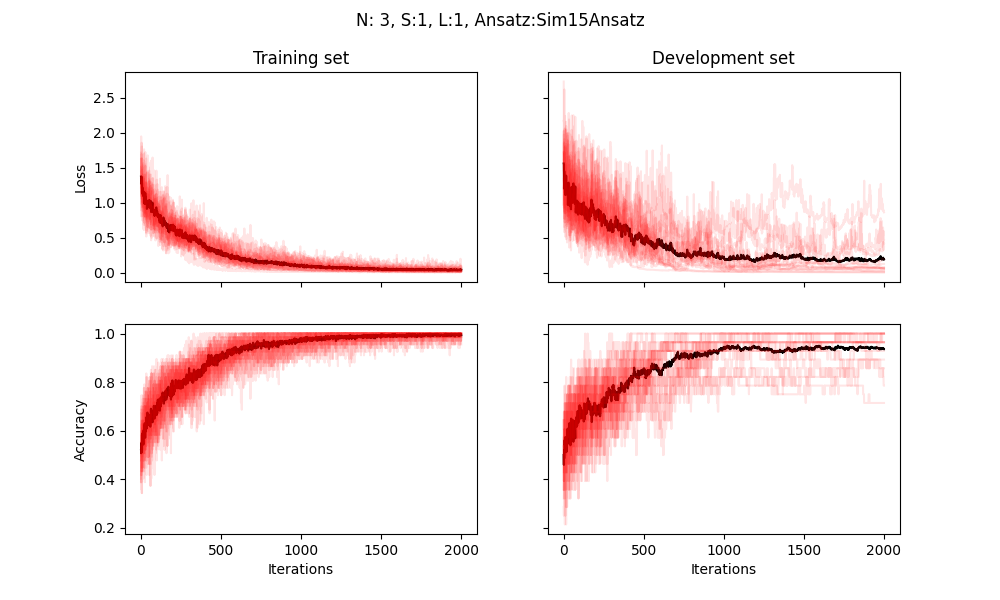}
\includegraphics[width=0.49\textwidth]{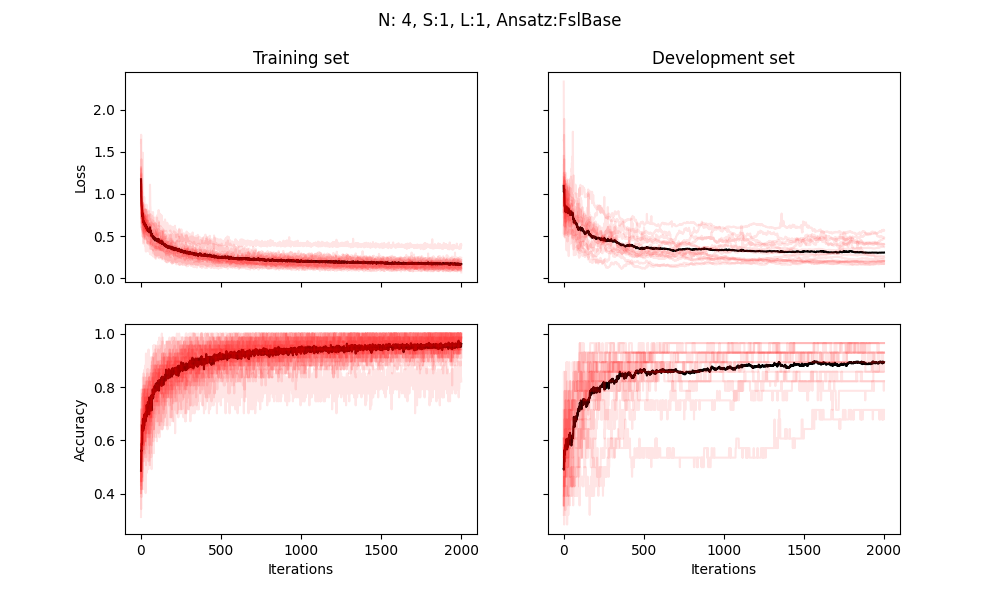}
\includegraphics[width=0.49\textwidth]{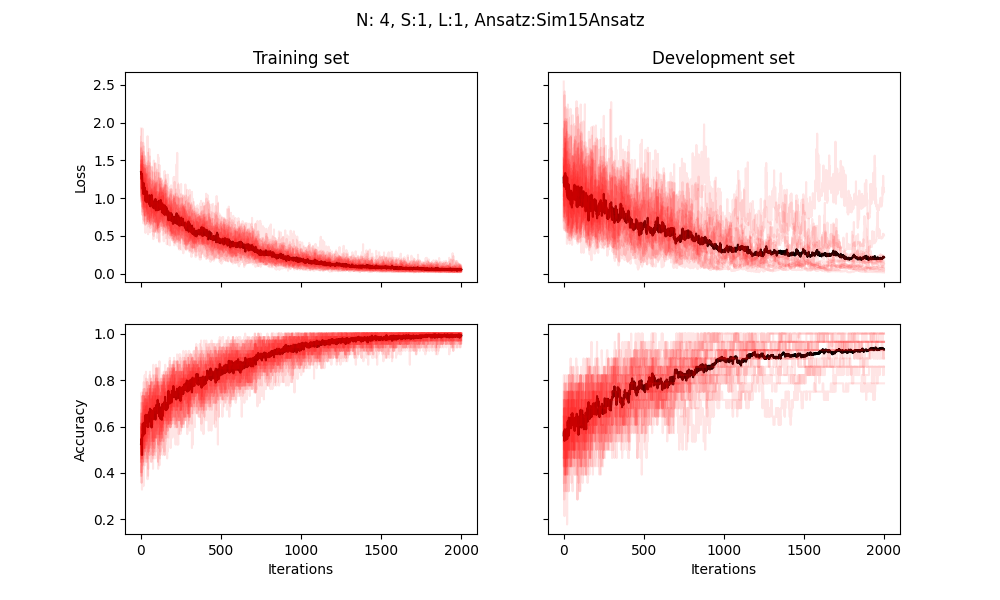}
\includegraphics[width=0.49\textwidth]{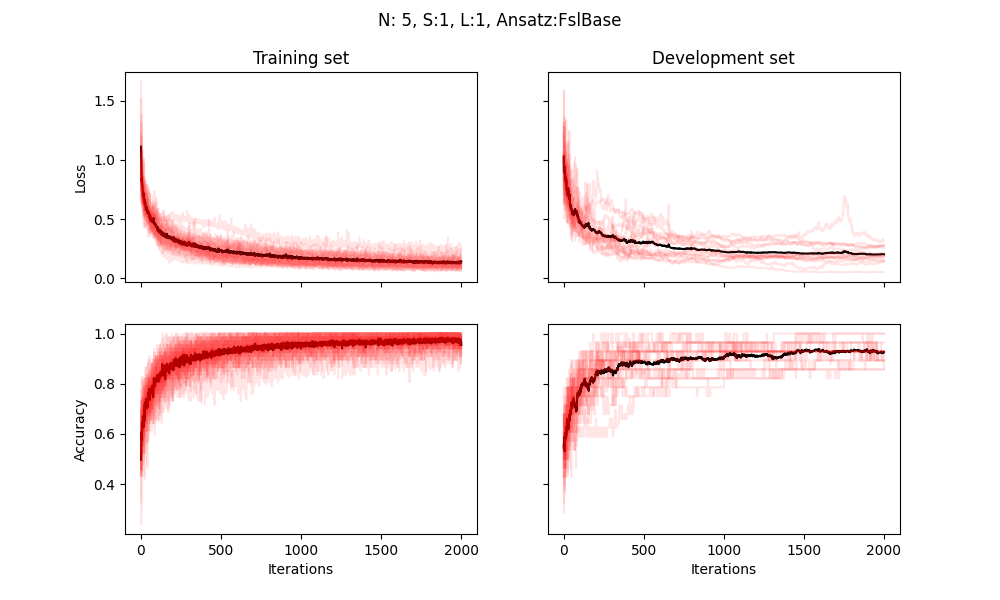}
\includegraphics[width=0.49\textwidth]{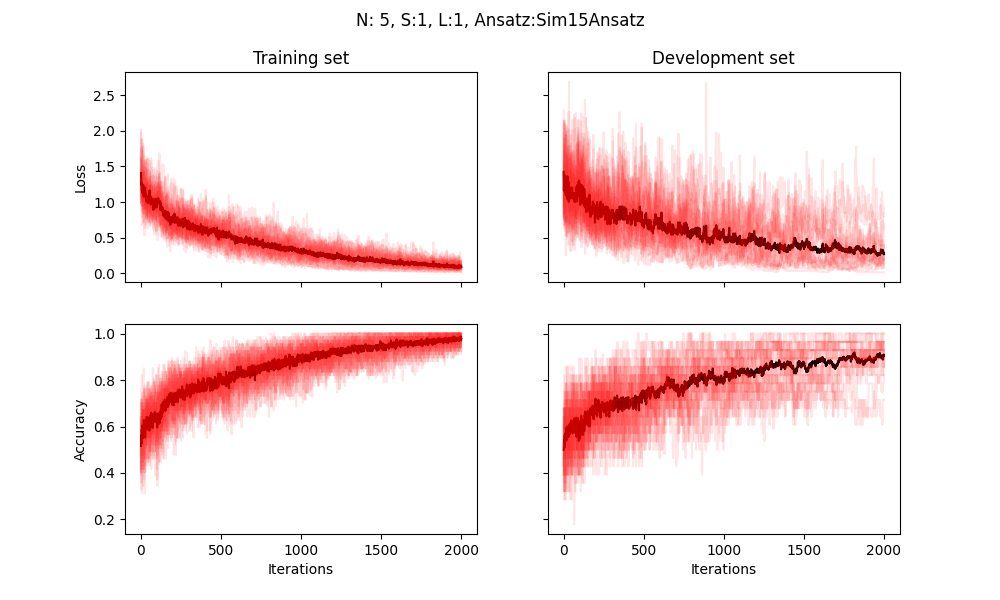}
\caption[Spread of \mya for small datasets]{Training spread for two types of \mya on the MC task after, average in black and each run in red. On the left, FSL base and on the right Sim15.}
\label{fig:testspread} 
\end{figure}

%% file: tables/accuracies.tex
\begin{table}[b]
\centering
\resizebox{\textwidth}{!}{%
\begin{tabular}{c|ccc|clr|crr|}
\cline{2-10}
\textbf{}                        & \multicolumn{3}{c|}{\textbf{Seen}}                                                              & \multicolumn{3}{c|}{\textbf{Mixed}}                                                                                    & \multicolumn{3}{c|}{\textbf{OOV}}                                                                                 \\ \hline
\multicolumn{1}{|c|}{\textbf{N}} & \multicolumn{1}{c|}{\textbf{FSL Base}} & \multicolumn{1}{c|}{\textbf{Fsl NN}} & \textbf{IQP}    & \multicolumn{1}{c|}{\textbf{FSL Base}} & \multicolumn{1}{c|}{\textbf{Fsl NN}} & \multicolumn{1}{c|}{\textbf{IQP}}      & \multicolumn{1}{c|}{\textbf{FSL Base}} & \multicolumn{1}{c|}{\textbf{Fsl NN}} & \multicolumn{1}{c|}{\textbf{IQP}} \\ \hline
\multicolumn{1}{|c|}{\textbf{2}} & \multicolumn{1}{c|}{0.7473}            & \multicolumn{1}{c|}{0.7185}          & \textbf{0.8306} & \multicolumn{1}{c|}{0.6338}            & \multicolumn{1}{l|}{0.5970}          & {\color[HTML]{3B3B3B} \textbf{0.6510}} & \multicolumn{1}{c|}{0.5361}            & \multicolumn{1}{r|}{\textbf{0.7260}} & {\color[HTML]{3B3B3B} 0.5288}     \\ \hline
\multicolumn{1}{|c|}{\textbf{3}} & \multicolumn{1}{c|}{\textbf{0.8949}}   & \multicolumn{1}{c|}{0.7311}          & 0.8521          & \multicolumn{1}{c|}{\textbf{0.6878}}   & \multicolumn{1}{l|}{0.6259}          & {\color[HTML]{3B3B3B} 0.6432}          & \multicolumn{1}{c|}{0.5269}            & \multicolumn{1}{r|}{\textbf{0.8356}} & {\color[HTML]{3B3B3B} 0.5215}     \\ \hline
\multicolumn{1}{|c|}{\textbf{4}} & \multicolumn{1}{c|}{0.7946}            & \multicolumn{1}{c|}{0.6866}          & \textbf{0.8610} & \multicolumn{1}{c|}{0.6363}            & \multicolumn{1}{l|}{0.5723}          & {\color[HTML]{3B3B3B} \textbf{0.6519}} & \multicolumn{1}{c|}{0.5516}            & \multicolumn{1}{r|}{\textbf{0.7534}} & {\color[HTML]{3B3B3B} 0.4945}     \\ \hline
\end{tabular}%
}
\caption{Accuracy table for FSL Base, FSL NN, and IQP \mya}
\label{tab:accuracies}
\end{table}

%% file: figures/2test.tex
\begin{figure}[ht]%
\centering
\includegraphics[width=0.49\textwidth]{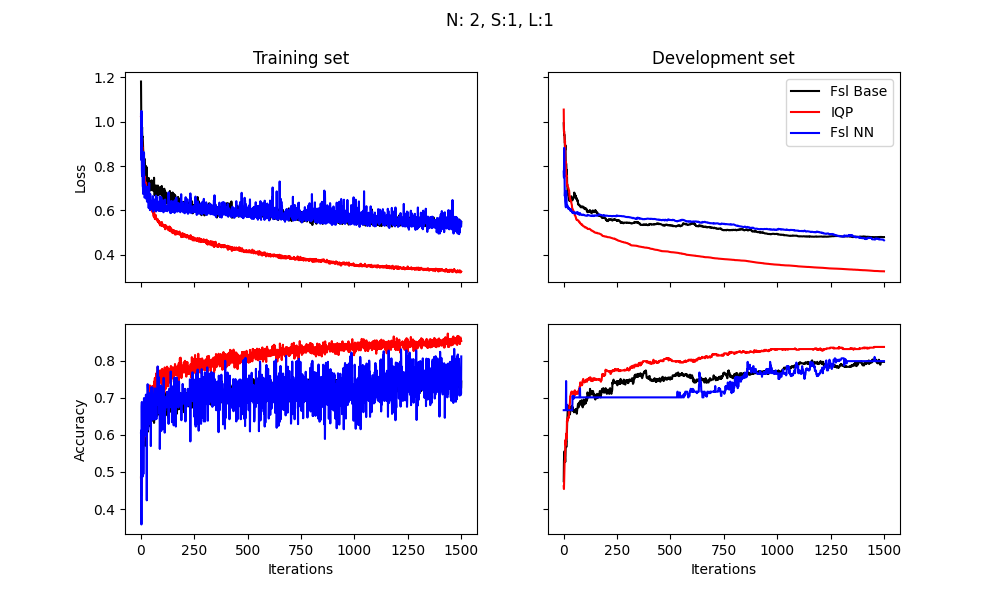}
\includegraphics[width=0.49\textwidth]{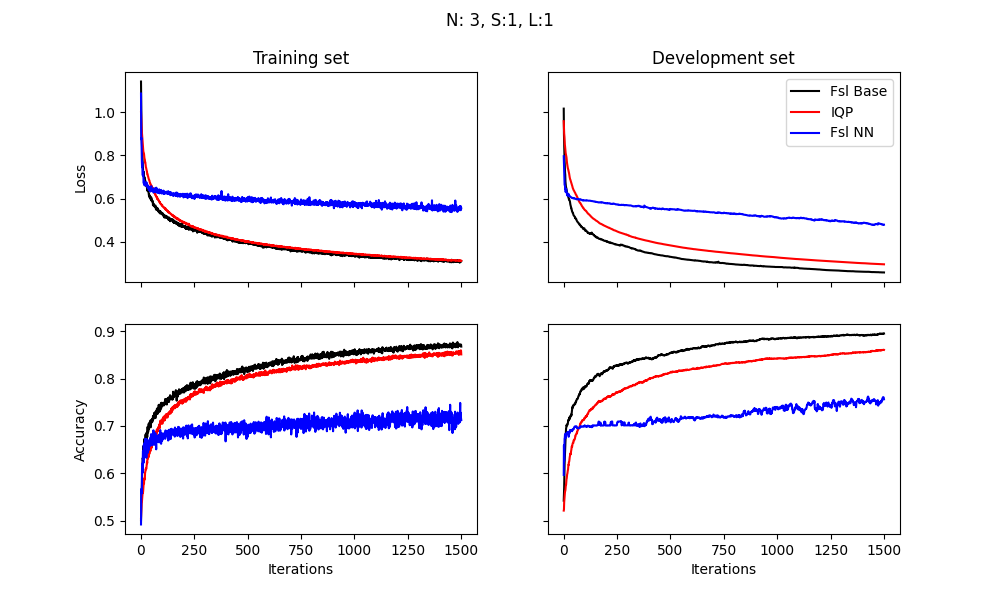}
\includegraphics[width=0.49\textwidth]{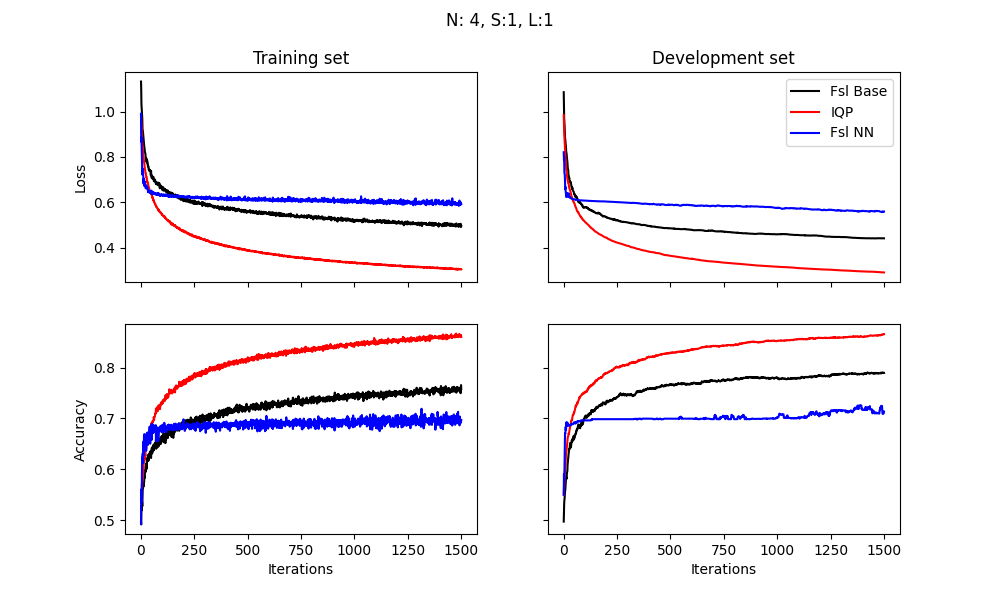}
\label{fig:2spread} 
\caption[Comparison between three \mya for bigger datasets]{Training behaviour comparison between FSL NN, FSL Naive and IQP.}
\end{figure}

%% file: 7_ResandCon.tex
\chapter{Conclussion, Discussion, and Further Work}\label{chap:conclusion}
\section{Conclusion}
After proposing a framework for applying Few Shot Learning to Quantum Natural Language Processing and testing it in various scenarios to learn about its behaviour, there are a few things we can ascertain.

First of all, we proposed two different types of PQE within the framework outlined that served two purposes: FSL Base tested a low classical cost implementation of FSL that seeks to minimally preserve some structure; FSL NN, on the other hand, seeks to establish a PQE that could improve the accuracy of the model with OOV words by relying more on the classical system.

Testing both of these PQEs we found out that this model of FSL exhibits faster convergence than traditional models albeit at a cost in accuracy. We also found out that the type of PQE chosen impacts the performance of the model for OOV words.

FSL NN has a higher classical computational cost than any of the other parametrisations, as a neural network has to be trained, but it does save in quantum computation time, which is the resource we are limited in and which we want to use less of.

Furthermore, this framework is extremely versatile, as we designed it to be extremely modular and adaptable to the characteristics of both the classical and quantum systems available. And, indeed, the testing done in this work implies that different versions of this framework can work for different use cases. As mentioned in the previous chapter, FSL NN is best suited for when OOV words are preferred, and FSL Base excels in fast convergence with minimal accuracy loss for sufficiently big training circuits. FSL Base also excels in the context where the unique word set has a large cardinality, but the set of valid sentences is not comparatively big, which traditional \mya would struggle.

As what happens with \mya usage, the optimal FSL PQE does not work with all implementations, a previous analysis of the situation is needed to first gauge what form of PQE would work best, what is the user willing to sacrifice to obtain the most of the quantum system, and what quantum resource is more scarce (e.g. the set of gates available, the amount of runs, the qubit connectivity, entanglement, etc.).

What underlies this method is also what grants its flexibility, the deterministic mapping between the classical embedding and the quantum parameters. This is not set concretely and the only tenant is the preservation of some structure that connects the word space structure with the task at hand. This also includes the choosing of the classical embeddings, as they are constructed to represent some type of structure between the words. Since DisCoCat has a heavy meaning component, the preservation of meaning through the inner product was the structure chosen in this work, but many others could be implemented.

Regardless, we showed that an FSL scheme can work to save quantum resources and further extract more power from quantum systems in the NISQ era.

\section{Discussion}

As it usually happens with works of the quantum type, this work was limited by the resources available. The task, dataset, and \mya chosen to test FSL had to be chosen in the context of maximising the resource allocation available to us. The training for a single model ran for about a week, not including the troubleshooting time. So even though higher circuit sizes are usually used in the literature, we had to restrict our testing space to smaller systems.

Since we try to leverage the vast amount of information already existing in classical NLP, the classical embeddings also play an important role, as the quality of the parameters obtained from the deterministic mapping from classical to quantum depends on the quality of the classical embeddings. Thus, the results are heavily influenced by the chosen embeddings and the manipulation of these vectors through the training process.

This framework is designed to work specifically with DisCoCat, other methods for QNLP could make use of the principles outlined in this work, but it would not be a one-to-one mapping and it is not something studied here.

This framework was also studied for the MC task. Many more tasks exist and could benefit from FSL, but due to the scope of this work being limited to the design and limited testing of the method, it is something that we leave to future work. However, within the framework of DisCoCat, adapting this to other tasks and sentence spaces of other sizes is straightforward and does not require work different from that done for the MC task.

Finally, since FSL involves a separate layer distinct from the one traditionally trained, an increased gate cost exists compared to traditional ans{\"a}tze. This cost needs to be taken into account if we want to implement this framework as part of the evaluation of resources needed to design an FSL PQE.

\section{Further Work}
As this work is the first proposal of this framework, more testing would need to be done on it. First, with higher computational resources, bigger circuit sizes and bigger datasets need to be tested how the generalisation ability present in classical FSL manifests itself in the quantum case. 

Other tasks need to be tested, including but not limited to  Pronoun Resolution, Paraphrasing, and Classifying to more than two labels. 

Testing also needs to be done to see the effect different PQE, different classical embeddings, and different mappings have on the performance of the model, both in terms of its convergence and its OOV performance. Classical embeddings in which different pregroup types correspond to different tensor ranks can also be studied.

Finally, even though AAE is not something paramount to our implementation of FSL, after reviewing the literature it does seem to be also a viable alternative to reduce quantum costs and should be researched. As of the time of writing, AAE has not been used in QNLP tasks.

\section{Final Thoughts}

We strove to design a framework for implementing FSL on quantum circuits based on the vast corpus of information that already exists on classical word embeddings. This method is flexible and was shown to work and, in certain cases, provide certain advantages compared to the traditional methodology. Even though further testing is necessary to see where and how this framework excels, we are confident in its ability to help further the uses of quantum systems in an era that, by its very nature, begs for some way to extract a lot from a few.

Working in the NISQ era is a burden for everybody, those who want to design algorithms and see them come to fruition, those who design and build the very quantum systems limited by the noisy and unpredictable nature of quantum theory in the non-quantum macroscopic world, those who try to develop useful ways to use NISQ systems to show that the quantum revolution is here and is just going to get better.

We hope that this work helps further the world of quantum systems and helps improve the way we conceive and use quantum devices. And we hope to one day see a world where NISQ is talked about like we talk about those big computers who used to occupy large rooms and made so much noise it felt as if they were alive. NISQ is not only marked by noise and problems but also by hope the universe gives us one more gift.

%% file: app_QuantumSupremacy.tex
\chapter{On Quantum Supremacy in the NISQ Era} \label{app:QuantumSip}

Quantum supremacy is a tricky thing to define. Broadly, we can define quantum supremacy as a point in development when a quantum computer can perform a task that a classical system cannot, effectively refuting the extended Church-Turing thesis \cite{harrow_quantum_2017}. The Church-Turing thesis states that we can efficiently simulate (i.e. in polynomial time) any model of computation on a turning machine  \cite{kliesch_dissipative_2011}. We can see glimpses of this in Feynman's original call to construct a quantum machine to simulate a reality that is quantum.

As mentioned in section \ref{sec:QPC,HPC}, quantum algorithms have been developed that can take advantage of the current state of computation. However, this does not necessarily satisfy the select few of the field that would be content only with a real implementation of quantum supremacy. But even though this has already been achieved in 2019 when a team from Google efficiently sampled a quantum circuit in 200 seconds when an equivalent classical problem would have taken 10,000 years \cite{arute_quantum_2019}, this is a toy problem designed to give the quantum computer the biggest advantage.

Even for practical problems in which quantum advantage could have been assumed to have been achieved, like Shor's factorization algorithm, this isn't the case, since the classical adversary is just the current state of the art, not a provably best-case scenario. The notion of advantage has to be altered from the proof-centric traditional mathematical sense, to be useful in the NISQ era. A more flexible definition could take into account not only the problems of interest but also what current classical and quantum devices could achieve.

To this end, reference \cite{hibat-allah_framework_2024} defines Practical Quantum Advantage (PQA) to be the ability of a quantum computer to perform some useful task in some way better than a classical processor could. This opens the definition to a lot of flexibility and qualitativeness and yet allows for practical demonstrations and proofs.

This definition is further subdivided into more specific terms, with the two most relevant for the conversations being Provable PQA (PrPQA) and Robust PQA (RPQA). The first one encapsulates the traditional notion of advantage, where a mathematical proof has to be given and so is robust and impervious to any improvement in the algorithms or the systems. RPQA, in turn, has under its umbrella most of the traditional notions of quantum supremacy, where a quantum system is currently outperforming its classical adversary, but this could change if a new classical algorithm is developed.

Quantum advantage, even in noisy systems, was demonstrated in 2017 \cite{riste_demonstration_2017}, and in 2022, researchers at Terra Quantum showed a quantum advantage in traditional ML tasks such as regression, optimisation, and classification, with the advantage coming in both speed and accuracy \cite{perelshtein_practical_2022}. However, these are just examples of RPQA, since they do not contain complexity arguments and are just comparing with the state of the art.

As with the original algorithms developed by Shor and Deutsch, most of the advantage in QML comes from using processes which are known the be advantageously run on quantum systems. For example, in \cite{yamasaki_advantage_2023} it was proven that a family of tasks can be constructed that can be run in polynomial time only on quantum systems.

Finally, it is important to briefly mention simulation in the NISQ era, as it is that objective which inspired Feynman in the first place. A review of the state of the art in \cite{daley_practical_2022} claims that current methods already show quantum advantage in analogue systems, where the quantum system is directly encoded in the quantum circuit.

The NISQ era brings with it the need to redefine traditional concepts of advantage to be able to gauge the improvement of quantum systems. Quantum systems are not the only entities susceptible to noise, as advantage and supremacy have also been changed in the search for improvement and, one day, complete quantum advantage.


%% file: app_bias.tex
\chapter{On Bias} \label{app:bias}

The widespread use of NLP in decision-making that can impact a person's life necessitates a deeper look into the inner workings of the models that are being used. To give an example, it was revealed in 2018 that Amazon had used an AI tool to help parse CVs for software development roles, and it had learned through its biased training to discriminate against women \cite{dastin_insight_2018}.

This becomes even more present with the rise of tools like ChatGPT and other LLMs that for the unaware public, might look like some sort of AGI instead of stochastic parrots \cite{bender_dangers_2021}. LLM tools, such as ChatGPT and LLAMA, exhibit discrimination that passes from the qualitative into a measurable bias against certain vulnerable groups in practical cases such as salary or price negotiation and skills assessment \cite{haim_whats_2024}.

Bias in NLP has been broadly studied \cite{lalor_benchmarking_2022,hutchinson_social_2020,garrido-munoz_survey_2021,sun_mitigating_2019},
and it can be categorised to come from five different sources \cite{hovy_five_2021}: the data, the annotation process, input representations, the model, and the research design.

Data could be the first thought of many when looking for biases in NLP, as humans are biased and most training is done with organic human data \cite{zhu_aligning_2015}. As human activity is biased, the models that employ them also become biased, as happened in the case of Amazon. Bias will disproportionately favour men, as many of its sources reflect that \cite{garimella_womens_2019}. The models reflect this disparity and can affect those underrepresented in the data \cite{hovy_demographic_2015}. Mitigation techniques usually require heavy input from a human to make sure the data is either augmented to make it more balanced \cite{webster_mind_2018} or taking out the identifying aspects of the dataset, such as gendered pronouns and names \cite{zhao_gender_2018}.

Label bias comes from the bias within annotators, which can unconsciously impart their societal norms on data which might not be so clearly cut labelled \cite{sap_risk_2019}. This can be mitigated by using multiple annotators \cite{hovy_learning_2013} and also using disagreement between annotators as part of the training process \cite{fornaciari_beyond_2021}.

Input representation harks back to section \ref{sec:embeddings}, where different words whose meanings are associated appear close together, so the biases manifest themselves through the proximity of words such as man and homemaker, and man and programmer \cite{bolukbasi_man_2016}. Debiasing usually relies on manually debiasing the embeddings themselves.

Model bias exists through the exact quality of the training of models, that is, models have a goal of minimising the cost function and they will do that, even if that represents exploiting statistical structures in the data that will result in bias. Countermeasures rely essentially upon having a set of metrics with which to train a model and being flexible enough to discard those that impart a bias on the model \cite{hovy_five_2021}, or even divorcing from the ideal of hard numerical metrics \cite{ribeiro_beyond_2020}.

Finally, design bias is an expression of the overwhelming amount of data in English, that limits the exposure of models to other languages and the cultural ideas expressed through them.

Bias is a problem in human societies and, as models are usually a reflection of them, it becomes present when applying any model to any task. Any solution will not be definitive, and will not be a panacea that can just be left to operate on itself. Debiasing models require researchers to take an active role and question what might be hidden behind their choices and the data they rely on.

%% file: app_aae.tex
\chapter{On Approximate Amplitude Encoding} \label{app:aae}

Another technique that, as of the moment of writing, has not been tried in QNLP and could be promising is approximate amplitude encoding.

A few techniques exist when trying to encode classical data into a quantum state: basis encoding \cite{li_efficient_2023}, angle encoding \cite{ovalle-magallanes_quantum_2023}, and ansatz encoding, to name a few.

Amplitude encoding maps some classical vector $\Vec{a}=[a_0,a_1,...,a_n]$ into the amplitudes of some quantum state: $|\Psi \rangle = a_0| 0 \rangle + a_1 | 1 \rangle +...+a_n|n\rangle$ \cite{larose_robust_2020}. For example, if we wanted to encode the vector $\Vec{a}=[\frac{1}{\sqrt{3}},0,\frac{1}{\sqrt{3}},-\frac{1}{\sqrt{3}}]$, the appropriate state would be $|\Psi \rangle = \frac{1}{\sqrt{3}}| 00 \rangle + 0 | 01 \rangle +\frac{1}{\sqrt{3}}|10\rangle-\frac{1}{\sqrt{3}}|11\rangle$. 

It is straightforward to see how amplitude encoding would translate to applications in QNLP. Obtaining the inner product of two classical vector embeddings would be equivalent to obtaining the projection of one word state into another. Considering that a quantum state can be completely determined by the vector containing the coefficients of its basis states, the tensor product between two states corresponds to the tensor product of these two amplitude vectors. So encoding the word embedding into the amplitude of the quantum state would be the natural way to encode the word states learned through classical DisCoCat.

However, this beckons the same age-old story about gate cost for ideal Unitary maps. Approximate Amplitude Encoding (AAE) was developed \cite{nakaji_approximate_2022} to variationally encode real-value vectors into the amplitude of a quantum state. Since the measurements of a quantum state can only be real and positive, the technique to encode a real value is not so straightforward.

The trick is to linearly split the data into its real and negative components $| \text{Data} \rangle = | \text{Data}^+ \rangle + | \text{Data}^- \rangle$, and defining the quantum state using an ancillary qubit
\begin{align}\label{eq:aaetrick}
    | \Psi \rangle = | \text{Data}^+ \rangle|0\rangle + | \text{Data}^- \rangle|1\rangle.
\end{align}
After applying a Hadamard transform on the ancillary qubit:
\begin{align}
    |\Psi \rangle  = \frac{| \text{Data}^+ \rangle - | \text{Data}^- \rangle}{\sqrt{2}}|0\rangle + \frac{| \text{Data}^+ \rangle + | \text{Data}^- \rangle}{\sqrt{2}}|1\rangle
\end{align}

Post-selecting the ancillary qubit to be $|1\rangle$, we can be sure the remaining state is the state with the negative values encoded in the amplitude.

This methodology can be extended to apply to variational algorithms to train PQCs to output the approximately encoded classical state to arbitrary accuracy.

Further refinement was done in \cite{mitsuda_approximate_2023} to allow for the encoding of complex data.

Considering that variationally using AAE to encode the word embedding would still be susceptible to the problems described in chapter \ref{chap:Problem}, most importantly the exclusion of OOV words. Here would lie another application for FSL in QNLP, since we could use FSL to extend the power to AAE to generalise to OOV words, and all the training could be directly imported from the classical version of DisCoCat.